\documentclass[twocolumn,epsf,prb,amsmath,amssymb,showpacs]{revtex4-1}
\usepackage{graphicx}
\usepackage{url}
\usepackage{amsmath}
\usepackage{amssymb}
\usepackage{xcolor}
\usepackage[T1]{fontenc}
\newcommand{\nup}[1]{\hat{n}_{#1\sigma}}
\newcommand{\ndw}[1]{\hat{n}_{#1 \bar{\sigma}}}

\newcommand{\ciup}[1]{\hat{c}_{#1 \sigma}}

\newcommand{\cidagup}[1]{{\hat{c}^\dag}_{#1 \sigma}}

\definecolor{darkblue}{rgb}{0.,0.,0.4}
\usepackage[pdftex,colorlinks=true,linkcolor=blue,citecolor=blue,urlcolor=blue,breaklinks]{
hyperref }

\begin{document}
\title{Non-linear response to electric field in extended Hubbard models}
\author{D. Nasr Esfahani}
\email{Davoud.NasrEsfahani@uantwerpen.be}
\author{L. Covaci}
\email{lucian@covaci.org}
\author{F. M. Peeters}
\email{Francois.Peeters@uantwerpen.be}
\affiliation{Departement Fysica, Universiteit Antwerpen,
Groenenborgerlaan 171, B-2020 Antwerpen, Belgium}

\begin{abstract}
The electric-field response of a one-dimensional ring of interacting fermions, where the
interactions are described by the extended Hubbard model, is investigated. By using an accurate
real-time propagation
scheme based on the Chebyshev expansion of the evolution operator, we uncover various non-linear
regimes for a range of interaction parameters that allows modeling of metallic and insulating
(either charge density wave or spin density wave insulators) rings. The metallic regime appears
at the phase boundary between the two insulating phases and provides the opportunity to describe
either weakly or strongly correlated metals. We find that the {\it fidelity susceptibility} of the
ground state as a function of magnetic flux piercing the ring provides a very good measure of the
short-time response. Even completely different interacting regimes behave in a similar manner at
short time-scales as long as the fidelity susceptibility is the same. Depending on the strength of
the electric field we find various types of responses: persistent currents in the insulating regime,
dissipative regime or damped Bloch-like oscillations with varying frequencies or even irregular
in nature. Furthermore, we also consider the dimerization of the ring and describe the response of a
correlated band insulator. In this case the distribution of the energy levels is more clustered and the
Bloch-like oscillations become even more irregular.
\end{abstract}
\maketitle
\section{Introduction}
The investigation of real time dynamics of a closed system consisting of interacting particles is important not only for the evaluation of experimentally relevant quantities, but also supplies reliable information about
the general properties of the Hamiltonian as long as one measures an appropriate set of observables throughout the propagation process\cite{RevModPhys.83.863}. This is of interest especially when the dimension of the Hilbert
space is very large and accessing the whole energy spectrum is not possible. There exist several approaches to face the problem of real time propagation of closed interacting systems. Among them are the numerically exact
polynomial expansions\cite{RevModPhys.78.275} or the approximate Lanczos propagation method\cite{hochbruck_krylov_1997}, the state of art time dependent density matrix renormalization group(tDMRG)\cite{PhysRevLett.93.207204}
and non-equilibrium dynamical mean field theory(nDMFT)\cite{PhysRevB.77.075109,PhysRevB.88.075135}. The common thread for all these methods is that it is not necessary to access the whole spectrum in order to evaluate time
dependent expectation values, hence this makes it feasible to investigate a large class of interacting systems.
 The special case of electric breakdown of 1D Mott insulators has been realized experimentally either with a strong electric field\cite{sawano_organic_2005,taguchi_dielectric_2000} or through photo-induced metal insulator
 transitions in pump probe experiments\cite{iwai_ultrafast_2003,okamoto_photoinduced_2007}. Further
interest was recently triggered by the realization of fermionic optical lattice experiments, where
the electric field effect on systems with designed interactions could be realized\cite{nature_optical,PhysRevLett.94.080403,bloch_ultracold_2005,esslinger_fermi-hubbard_2010}.

 There exist exist several theoretical investigations on the real time dynamics of the Hubbard Hamiltonian, part of which focused on real time quench dynamics\cite{eckstein_interaction_2010, eckstein_new_2009,PhysRevLett.98.180601,genway_thermalization_2012,PhysRevA.82.011604},
 real time studies based on the relaxation dynamics of specifically prepared exited states\cite{PhysRevLett.100.166403} as well as the effect of an external electric field\cite{PhysRevLett.107.126601,PhysRevLett.106.196401,PhysRevB.77.075109,eckstein_damping_2011,PhysRevLett.105.146404,PhysRevLett.109.197401,takahashi_photoinduced_2008,mierzejewski_current_2010}.
 The electric break down of a one-dimensional Mott insulator has been theoretically investigated\cite{oka_breakdown_2003,PhysRevLett.95.137601} and the analysis was based on a Landau-Zener(LZ)\cite{Landau,Zener} mechanism,
 which showed an exponential decay of the probability
 of the initial ground-state as function of time in short time scales. The decay rate is a function of an exponential function with an exponent proportional to square of the charge
gap of the system\cite{PhysRevLett.95.137601}, however this is not universal and the dependence of the exponent on the charge
gap could deviate from quadratic type for specific cases\cite{PhysRevLett.108.196401}. We found there are situations
in which the breakdown is not simultaneous with the overlap of ground-state with only the first
excited state but also with higher energy states. This happens especially for
  insulating systems with larger charge gaps. This therefore makes inappropriate the use of a simple
two level approximation and the LZ parameter as a basis for comparing different
insulating systems. In order to alleviate these discrepancies of the two level approximation we
employ the recently proposed {\it fidelity susceptibility}\cite{you_fidelity_2007} as a
measure for the change of basis-set as function external field.
  This quantity is unbiased and can be calculated numerically exact. Throughout this work we use it
as a basis for comparing the response of different insulating systems to a constant electric field.

Beyond the short time-scale ground-state decay, a question that grasped the attention is how much
does the electric field response at longer time scales depends on ground-state properties
and/or interaction parameters.
A notable phenomenon that definitely depends on longer time scales and is beyond the ground state
decay mechanism based on the Landau-Zener(LZ) tunneling is the appearance of Bloch oscillations(BO).
The existence of Bloch oscillations has already been proven experimentally in semiconductor
super-lattices\cite{PhysRevB.46.7252,Leo1992943,PhysRevLett.70.3319,PhysRevLett.61.1639}.
Furthermore,
the damping of Bloch oscillations in a closed interacting system subjected to an uniform electric field has been
described theoretically within the Falikov-Kimbal model\cite{PhysRevB.77.075109}, the one-dimensional Hubbard spin-less model
\cite{PhysRevLett.105.186405}, where it is shown an integrable system shows current oscillations
with frequencies smaller than the normal BO when subjected to uniform weak field, and
in the one dimensional Holstein model\cite{vidmar_nonequilibrium_2011}, where authors report the
presence of an stationary state which carries a finite current.
Furthermore BO oscillations in electric break down of a 3-dimensional Hubbard model\cite{PhysRevLett.105.146404} is investigated. By using an extended Hubbard model one has
 the opportunity to design the interaction parameters in order to have better understanding about the mechanism of the formations of BO in different regimes, and it is the aim of
 this paper to investigate the differences between the non-linear response of different kinds of closed systems of interacting fermions both in the insulating and
the metallic regimes. We achieve this by employing a real time propagation scheme together with the
ground-state and spectral analysis. Based on our analysis it appear to be impossible for a closed
system to have an stationary state which carries finite stationary current. Our paper is organized
as follows: in Sec.~II we present our
model under study together with a brief description of the theoretical and numerical schemes. In Sec.~III(A) we present our analysis of the response to constant electric field for a system
of weakly interacting fermions, while in Sec.~III(B) we perform the same study but for strongly interacting fermions. Finally, in Sec.~IV we give our conclusions.

\section{Model and Method}
Our model under investigation is a one-dimensional closed system of interacting charged fermions with periodic boundary conditions. It can be described in the second-quantization formalism by an extended Hubbard model as follows:
\begin{eqnarray}\label{eq1}
\hat{H}&=& -\sum_{\langle ij \rangle \sigma} [h_{ij}(t) \cidagup{i}\ciup{j}+h.c.] + \sum_{\langle ij \rangle} \frac{1}{2}V_{ij}\hat{n}_i\hat{n}_j\nonumber \\
       &+& \sum_i U \nup{i}\ndw{i},
\end{eqnarray}
where $\langle .. \rangle$ represents the summation over the nearest neighbor sites. $\cidagup{i}$
and $\ciup{j}$ are the creation and annihilation fermion operators. The fermion density is defined
as usual as $\hat{n}_i=\hat{n}_{i\uparrow} + \hat{n}_{i\downarrow}$ with
$\hat{n}_{i\sigma}=\cidagup{i}\ciup{i}$. The first term in Eq.~(\ref{eq1}) represents the kinetic
energy, where the hopping amplitude is taken to be time-dependent and by using the Peierls
substitution becomes $h_{ij}(t)=h_{ij}(0)e^{\frac{ie}{\hbar c}\phi(t)}$ with $h_{ij}(0)=[h_0+(-1)^i\eta]$.
$\phi=\phi_{tot}/L$ is the total magnetic flux piercing the ring divided by the number of sites and $\eta$
models a dimerization term. Hereafter we consider $\hbar=e=a=1$, where $a$ is lattice constant.
Interactions are either local between fermions with opposite spins, described by $U$, or non-local
between fermions sitting on neighboring sites, described by $V_{ij}$. All of the coupling constants which are
 reported in the following are scaled with $h_0=1$. Throughout this work we consider an electric field,
 which is given by the time derivative of the flux, $\tilde{F}=-\dot{\phi}(t)$.

Starting from parameters at $t=0$ we find the ground state of the resulting Hamiltonian and
propagate it while considering the change of the coupling parameters as function of time. To find
the solution of the time-dependent Schr\"odinger equation,
\begin{equation}
 H(\phi(t))|\psi(t)\rangle = i\dot{|\psi(t)\rangle},
\end{equation}
one may write it as a superposition of the instantaneous eigenstates of the time-dependent
Hamiltonian as,
\begin{equation}\label{eqn1}
 |\psi(t)\rangle = \sum_n c_n(t)|n_{\phi}(t)\rangle,
\end{equation}
where $|n_\phi(t)\rangle$ are the instantaneous eigenstates of $H(\phi(t))$ with
$H(\phi(t))|n_\phi(t)\rangle=E_n(t)|n_\phi(t)\rangle$. By substituting $|\psi(t)\rangle$ as
expressed by Eq.~(\ref{eqn1}) into the Schr\"odinger equation and by using the change of variables
as
 $\tilde{c}_n(t)=c_n(t)e^{i\theta_{mn}(t)}$, one obtains the following set of coupled differential equations for the coefficients $\tilde{c}_n(t)$,
  \begin{equation}\label{eq2}
   \dot{\tilde{c}}_n(t) = -\sum_{m\neq n} e^{i\theta_{mn}(t)} \tilde{c}_m(t) \langle n_\phi(t)|\dot{m}_\phi(t)\rangle,
  \end{equation}
where $\theta_{nm}(t)= \int_0^t(E_n(\tau)-E_m(\tau))d\tau - i\int_0^t (\langle n_\phi(\tau)|\dot{n}_\phi(\tau)\rangle-\langle m_\phi(\tau)|\dot{m}_\phi(\tau)\rangle)d\tau$,
this change of variables is in fact a gauge transformation because $\theta_{nm}(t)$ are purely
real\cite{PhysRevLett.108.080404}.

The change of basis set as function of time manifests itself in the $ \langle n_\phi(t)|\dot{m}_\phi(t)\rangle $ term in the
right-hand side of Eq.~(\ref{eq2}). By starting from an eigenstate
of  the Hamiltonian at $t=0$ with $|c_n(0)|=1$, as long as the terms $\langle n_\phi(t)|\dot{m}_\phi(t)\rangle\simeq 0$ during the evolution, then one arrives
at the adiabatic regime where $|\psi(t)\rangle$ only follows the eigenstate of the instantaneous Hamiltonian and the coefficients $|c_n(t)|=1$ only consist of a phase that
is a combination of a geometrical Berry and a dynamical phase. For the non adiabatic regime,
Eq.~(\ref{eq2}) not only ensures the change in the magnitude of
$c_n(t)$ but each coefficient further accumulates a complicated phase consisting of dynamical and Berry phases produced by the other states.

If we consider the ground-state as the starting state for the time evolution, the quantity that
measures
the change of basis set as function of the external parameter
$\phi$ is the {\it ground-state fidelity}\cite{PhysRevE.74.031123} which is defined as
\begin{equation}
\varXi(\phi)=|\langle \psi_0(\phi)|\psi_0(\phi+\delta \phi)\rangle|.
\end{equation}

 By using perturbative arguments it is possible to see that there is a close relationship between
the ground-state fidelity and the coefficients that appear in
Eq.~(\ref{eq2}),
\begin{equation}
\langle n(t)|\dot{m(t)}\rangle=\dot{\phi}\frac{\langle
n_\phi(t)|\partial_\phi{H(\phi(t))}|m_\phi(t)\rangle}{(E_n-E_m)}.
\end{equation}
Therefore the change in the ground-state wave-function under an infinitesimal change of flux can
be written as:
\begin{equation}
 |\psi_0(\phi+\delta \phi)\rangle = \Lambda\left[|\psi_0(\phi)\rangle +\delta\phi \sum_{n\neq
0}\frac{\langle n_\phi|\partial_\phi{H(\phi)}|\psi_0(\phi)\rangle}{E_0-E_n}|n_\phi\rangle\right],
\end{equation}
where $\Lambda$ is a normalization factor. After normalization and considering $\delta\phi<<1$  one
obtains that
\begin{equation}
 |\langle \psi_0(\phi)|\psi_0(\phi+\delta \phi)\rangle|^2 = 1 - (\delta\phi)^2
\chi_\varXi(\phi),
\end{equation}
where $\chi_\varXi(\phi)$ is the {\it fidelity susceptibility} which is defined
as\cite{you_fidelity_2007,PhysRevB.88.195101},
\begin{equation}\label{eq3}
 \chi_\varXi(\phi) = \frac{1-{\varXi^2(\phi)}}{{(\delta\phi)}^2}= \sum_{n\neq 0} \frac{\langle
\psi_0(\phi)|\partial_\phi{H(\phi)}|n_\phi)\rangle^2}{(E_0-E_n)^2}.
\end{equation}

The leading term in the fidelity expansion is of the order of $\delta \phi^2$. When comparing the
terms in the right-hand side of Eq.~(\ref{eq3}) with terms
that appear in the right-hand side of Eq.~(\ref{eq2}) one may infer that a larger
$\chi_\varXi(\phi)$ leads to a more non-adiabatic character of the transition
due to the driving of the system by an external electric field. We will use the ground-state
fidelity susceptibility in the following sections as a basis for
the comparison of the short term response of different kinds of interacting fermions modeled by
Eq.~(\ref{eq1}). We do this in particular when the system is subjected to a constant and uniform
electric field.

Although the instantaneous eigenstate representation of the time-dependent Schr\"odinger
equation is very insightful, the solution of Eqs.~(\ref{eq2}) is
either very difficult or outright impossible for systems where the Hilbert space is very large
and having the eigenstates at each moment is very computationally expensive. For the case
 of interacting fermions with spin the dimension of the Hilbert space for a small system which
consists only 10 sites at half filling is $\sim 63000$, which makes solving Eqs.~(\ref{eq2})
  almost impossible.

An alternate way to deal with the time-dependent Schr\"odinger equation is to exploit the
form of the unitary time evolution
  operator:
\begin{equation}
  \hat{U}(t)=\mathcal{T}e^{-i\int_0^{t_f} \hat{H}(\tau)d\tau}\simeq\prod_k^N e^{-i\hat{H}(t_k)\delta
t},
\end{equation}
where $\delta t=t_f/N$. Therefore, the problem is
  reduced to a stepwise change of the Hamiltonian and relaxation of the system with a time step
equal to $\delta t$. Over each time-step the Hamiltonian is considered to be time-independent and
the relaxation of the wave function can be easily performed, by using the Chebyshev
propagation method\cite{RevModPhys.83.863}, which considers an expansion of the evolution operator. The
wave-function at $t_i+\delta t$ can now be written as:
\begin{eqnarray}
|\psi(t_k+\delta t)\rangle &=& e^{-ib\delta t}[ J_0(a\delta t)I \nonumber \\
                       &+& \sum_{s=1}^{\infty} 2{(-i)}^s J_s(a\delta t)
T_s({\tilde{H}})]|\psi(t_k)\rangle,
\label{cheby}
\end{eqnarray}
where $\tilde{H} = (\hat{H}-bI)/a$ with $b=(E_{max}+E_{min})/{2}$ and $a =
(E_{max}-E_{min})/(2-\epsilon)$. $J_s$ are $s$-th order Bessel functions of the first kind
and T$_s(x)$ are the Chebyshev polynomials which obey the recursion relation, $T_s(x) =
2xT_{s-1}(x)-T_{s-2}(x)$.
 $\epsilon$ is introduced in order to make sure that the absolute value of the extreme eigenvalues
of $\tilde{H}$ is less than 1. This is crucial for the Chebyshev method because the arguments
of Chebyshev polynomials accept only values in the interval $[-1,1]$. We
truncate the series in Eq.~(\ref{cheby}) such that the propagated wave function becomes normalized
up to machine accuracy in order to reduce error accumulation
  during the stepwise propagations. Moreover this also ensures that the propagation operator is
unitary up to machine accuracy. Having the wave function at each
  time-step, then the coefficients from Eq.~(\ref{eq1}), $c_n(t)=\langle n_\phi(t)|\psi(t)\rangle$,
could be calculated for analysis purposes only whenever it is necessarily or possible to do so.

  In order to have some insight about the nature of the wave-function, $|\psi(t)\rangle$, we
further calculate the structure factors that are defined as,
 \begin{equation}
  C_{X}(q) = \frac{2}{L^2}\sum_{i=1}^L\sum_{j=1}^{L/2} e^{i qr_{i,i+j}} \bar{X}_{i,i+j}
 \end{equation}
 where
$\bar{X}_{s,k}=\langle\hat{X}_s\hat{X}_k\rangle-\langle\hat{X}_s\rangle\langle\hat{X}_k\rangle$, $s$
and $k$ are the site indices (summation over $L/2$ for $j$ is introduce because $\bar{X}_{s,k}$ is symmetric around $\bar{X}_{s,s+L/2}$ due to
the periodic boundary condition we considered)and $r_{s,k}$ is the distance between site $s$ and
site $k$. We report
spin density wave(SDW) order parameter $O_{SDW}=C_{\hat{S}}(\pi)$ with
$\hat{S}_s=1/2(\nup{s}-\ndw{s})$ and charge density wave(CDW) as $O_{CDW}=C_{\hat{n}}(\pi)$,
 where $\hat{n}_s=\nup{s}+\ndw{s}$ is the local density operator. We also report the value of
the current as function of time, which is defined as the expectation value of the current operator,
$\hat{J}=\frac{i}{L} \sum_{\langle sk \rangle \sigma} [h_{sk}(t) \cidagup{s}\ciup{k} - h.c.]$.

\section{Results}
In the following we set $h(0)=h_0=1$ and all the coupling constants are scaled with $h_0$.
Moreover we define the uniform electric field, $\tilde{F}$, as $\phi(t)=-\tilde{F} t$. For the sake of simplicity we
define $F=\tilde{F}/2\pi$. We consider the time steps to be $\delta t = 0.005$. We have
tested all the results against a finer time grid in order to ensure that there is no
quantitative difference over the parameter
range considered here.

\begin{figure}[ttt]
\begin{center}
   \includegraphics[width=\columnwidth]{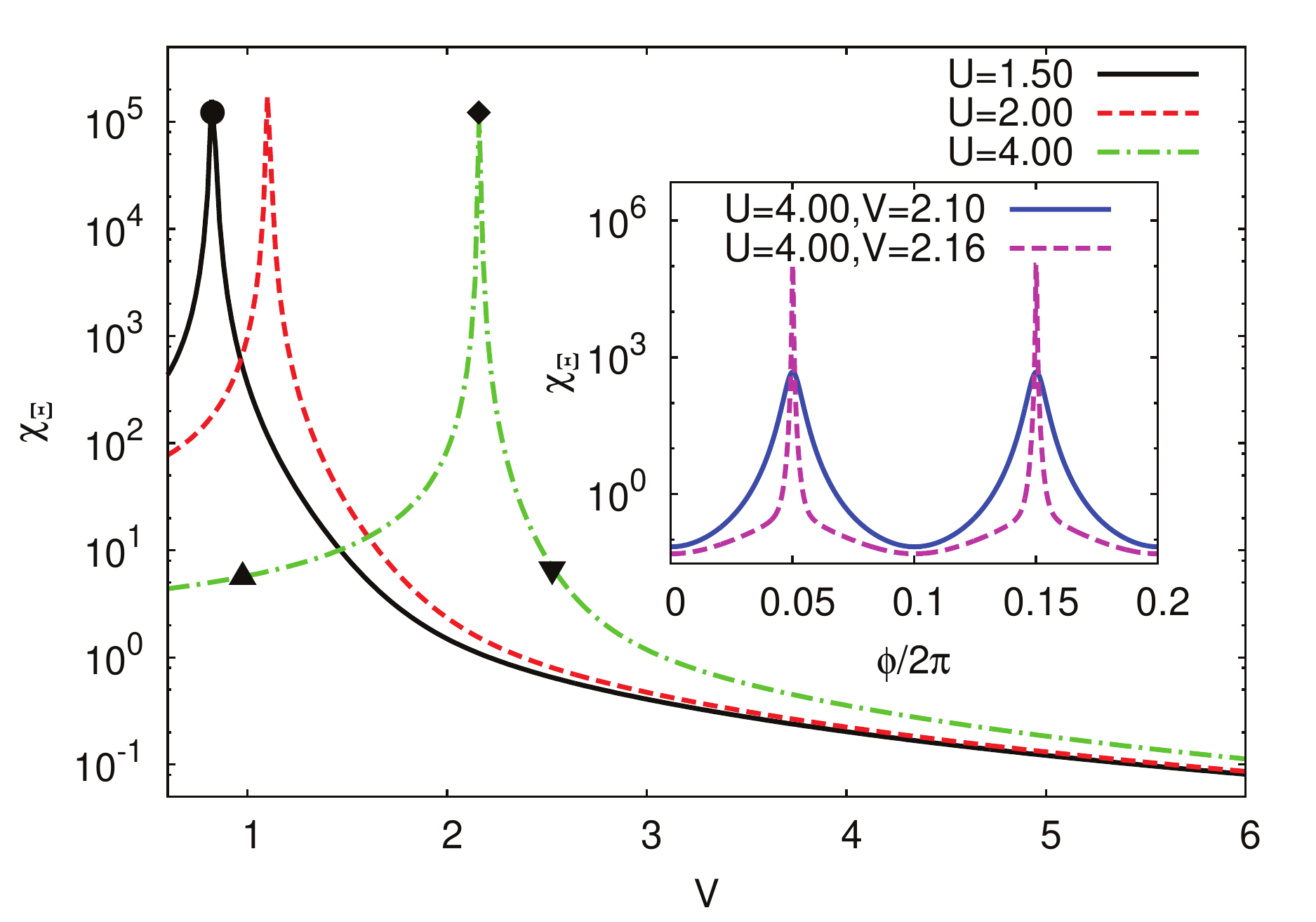}
   \caption {(Color online) Fidelity susceptibility for a ring with $L_{sites}=10$ and
$N_\uparrow=N_\downarrow=5$
at $\phi=0.1\pi$ as function of interactions. The inset shows the
   fidelity susceptibility as function of $\phi/ 2\pi$ for different sets of parameters. The points
represent specifically chosen pairs of parameters $U,V$ in order to model a weakly interacting
metal (circle), a SDW insulator (triangle), a CDW insulator (upside down triangle) and a strongly
interacting metal (diamond).}
   \label{fig1}
   \vspace{-0.5cm}
\end{center}
\end{figure}

We start by showing in Fig.~\ref{fig1} the fidelity susceptibility, $\chi_\varXi(\phi)$, at
$\phi=0.1\pi$ for a system consisting of 10 sites at half-filling for different values of $U$ and as
a function of $V$, we use $\delta \phi=10^{-3}$ for the calculations presented in Fig.~\ref{fig1}. As is clear from the inset of Fig.~\ref{fig1}, $\chi_\varXi(\phi)$ acquires
the largest value at $\phi_{anti}=0.1\pi$, which is an anti-crossing point between the
ground-state and an excited state. Notice that here we calculate $\chi_\varXi(\phi)$ numerically
exact with the use of the Lanczos method and do
not use the perturbative form introduced in Eq.~(\ref{eq3}).

Notice that the susceptibility is largest, almost diverging, at specific values of $V$ for each $U$, whenever the relation $U\simeq
2V$ is satisfied. This relation represents the boundary which separates
the SDW and
CDW phases\cite{PhysRevLett.99.216403}, and was obtained within the DMRG approach for 1D chains of
larger dimensions. However, it is obvious that $\chi_\varXi(\phi_{anti})$ can provide a good
estimate on the location of the SDW-CDW phase boundary, although it does not
provide any information about the details of the wave-function (whether it describes SDW or CDW).

  In order to compare the non-linear response of different kinds of interacting systems we
analyze different sets of interaction and hopping parameters.
   In particular we study three different cases: first we consider a system with
$U=1.5$ and $V=0.82$, marked with a circle in Fig.~\ref{fig1}, which shows an almost diverging
   $\chi_\varXi(\phi_{anti})$ and has
a vanishingly small charge gap, $\Delta_{charge}(\phi_{anti}) \simeq 10^{-3}$,
   and therefore could be considered as a {\it weakly interacting metal}. Secondly, we use a
dimerization parameter $\eta=0.4$, which opens up a gap and the system behaves as a {\it correlated
band insulator} (BI). Finally, we choose a stronger interacting system with $U=4.0$ and three
   different values of $V=$0.94, 2.56 and 1.16. Two values, $V=0.94$ (a {\it SDW insulator}, marked
with a triangle in Fig.~\ref{fig1}, $\Delta_{charge}(\phi_{anti})=1.44$) and $V=2.56$ (a {\it CDW insulator}, marked with an upside down
triangle in Fig.~\ref{fig1}, $\Delta_{charge}(\phi_{anti})=1.36$) are chosen such that $\chi_\varXi(\phi_{anti})$ is the same.
   We also consider $V=1.16$ on the phase boundary between SDW and CDW with an almost diverging
$\chi_\varXi(\phi_{anti})$ (marked with a diamond in Fig.~\ref{fig1}).
    The latter case also has a vanishingly small charge gap but it
should be considered as a {\it strongly interacting metal}.

\subsection{Weakly interacting system}
\begin{figure}[ttt]
\begin{center}
   \includegraphics[width=\columnwidth]{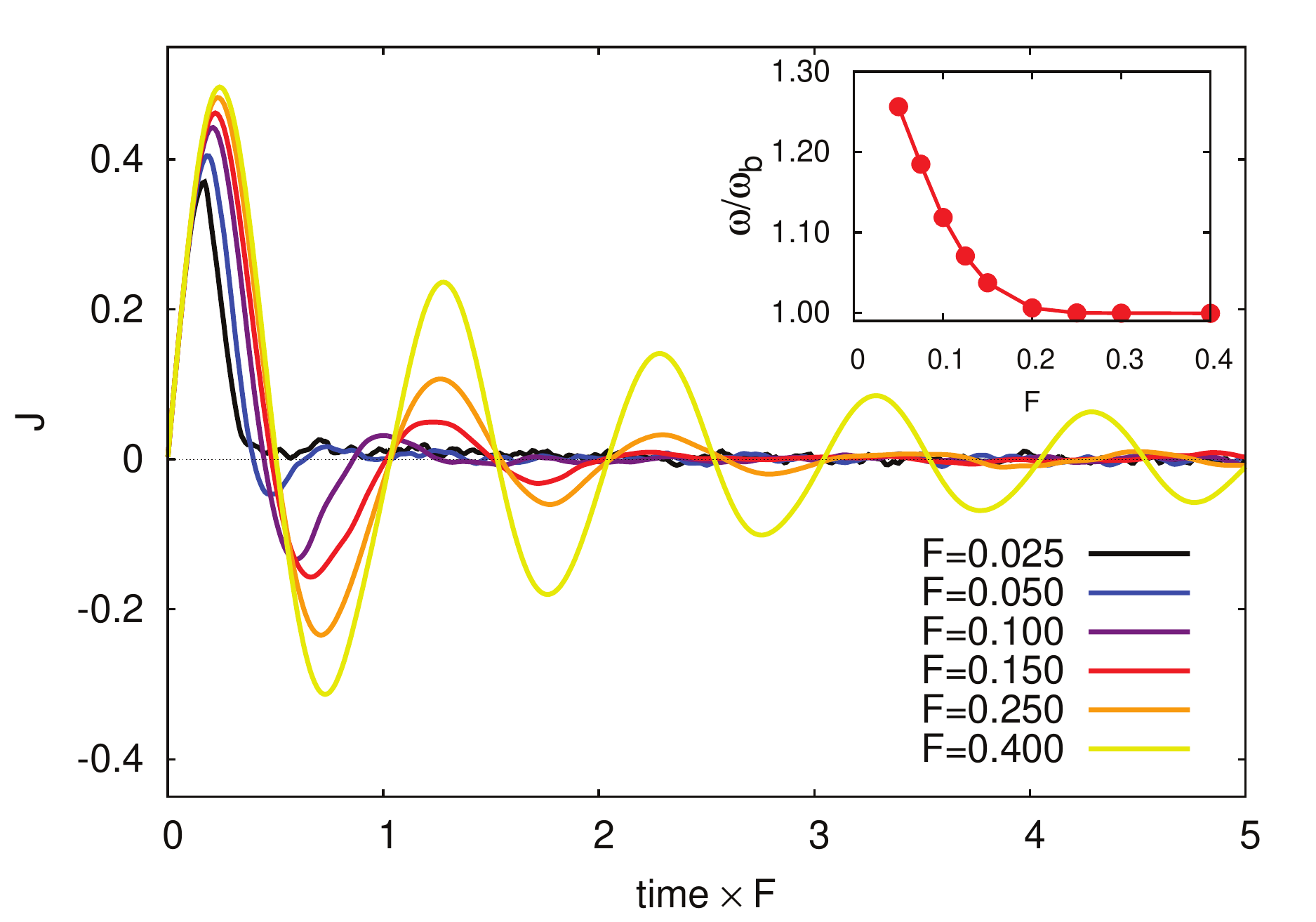}
   \caption {(Color online)  Current as function of time for a ring with $U=1.5$, $V=0.82$,
$L_{sites}=10$, $N_\uparrow=N_\downarrow=5$ and for different electric
   field strengths. Inset: the frequency of the BO for different electric fields and the same
parameters of the main graph with $\omega_b=F$.}
   \label{fig2}
   \vspace{-0.5cm}
\end{center}
\end{figure}

In Fig.~\ref{fig2} we show the current as function of time for a system with $U=1.5$ and $V=0.82$ for different electric field strengths. For illustrative purposes we start the analysis of the graph from
 the largest field, $F=0.4$, where it shows a regular damped BO in the time domain of interest. As we stated previously, $\chi_\varXi(\phi_{anti})$ is largest
 at the anti-crossings, thus the probability transfer from the ground-state to excited states (also
in analogy with LZ theory) is enhanced. Therefore at each
 anti-crossing there is a high probability of transfer from a right going wave ($-\partial E_n(\phi)/\partial\phi>0$) to another right going wave. When the field is
 strong enough this transfer is very efficient such that the wave-function has a significant overlap
with {\it only one} of the eigenstates of the
 instantaneous Hamiltonian. Finally when the maximum energy is reached, the wave-function will start
having significant overlap with left-going states and the current will change sign.
 This reflection for the high field case happens exactly at $t=(2F)^{-1}$.

\begin{figure}[ttt]
\begin{center}
   \includegraphics[width=\columnwidth]{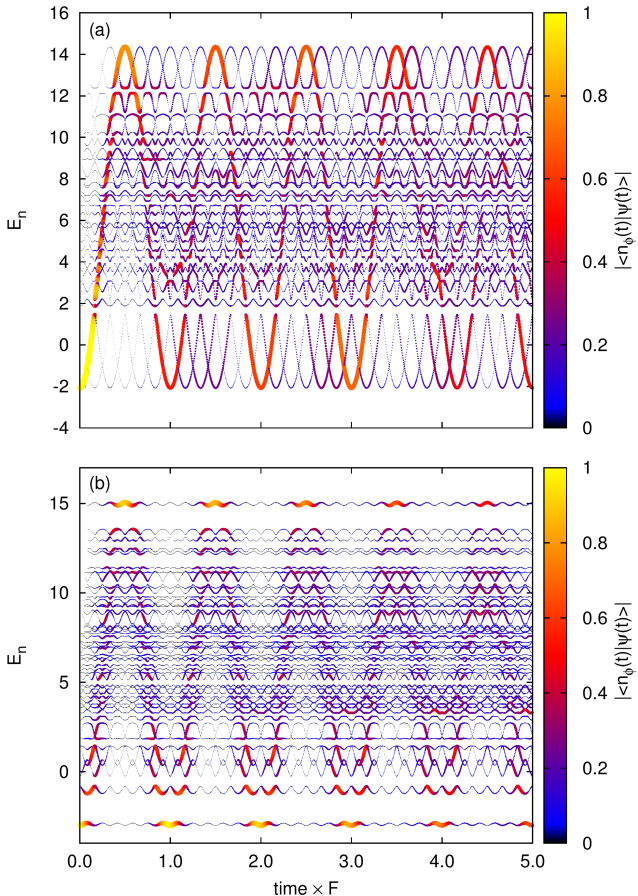}
   \caption {(Color online) (a) Eigenvalues of the instantaneous Hamiltonian as function of time
for a system with L=6 at half filling, U=1.5,V=0.82, F=0.4. The colors and the size of the points
are given by the overlap of the time-dependent wave-function with the instantaneous eigenstates,
$|\langle n_\phi(t)|\psi(t)\rangle|$; (b) The same as (a) but with a dimerization
    parameter $\eta=0.4$ and F=4.0.}
   \label{fig3}
   \vspace{-0.5cm}
\end{center}
\end{figure}

 To better understand the above description of the large field response, we plot in Fig.~\ref{fig3}(a)
 the eigenstates of the instantaneous Hamiltonian as a function of time for a smaller system, with
$L=6$ at half-filling, for $F=0.4$ and the same interaction parameters. Both the size of the points
and their color code represent the magnitude of the overlap of the time-dependent wave-function with
the instantaneous eigenstates of $\hat{H}(t)$. Note that the spectrum is periodic with $2\pi/L$,
thus the first anti-crossing happens at $tF=0.5/L=0.833$.
  This smaller ring shows very similar behavior to the one presented in Fig.~\ref{fig2} when subjected to a strong field, except that the magnitude of the current is smaller. The formation of a coherent path for
   the probability transfer throughout the spectrum and the reflection at the topmost state when $t=(2F)^{-1}$ can be clearly seen. However, a dissipative loss of the probability to both left-going and right-going waves is possible and the current becomes damped as function of time. For higher fields the probability transfer is more efficient, which means that the damping of BO is suppressed.

{\it Weak fields.} Looking back to Fig.~\ref{fig2}, the weakest field response, for $F=0.025$, is comprised of two non-linear effects. First, the state with high probability is reflected sooner, well
before it arrives at the other edge of the spectrum. This could be inferred from the fact that the current changes sign sooner than in the high field case. Second, when the field is weak the probability
transfer to excited states is smaller, which means that at each higher energy anti-crossing there is a finite probability of remaining in the state with lower energy, which will contribute with a negative
sign to the total current. Therefore after an initial increase in current, the wave-function will
overlap with equally right-going and left-going instantaneous states and one ends up with a
quasi-stationary
regime in which the current is very small and fluctuates around zero.

We further elucidate this behavior by expressing the current as function of instantaneous
eigenstates of $\hat{H}(t)$,
  \begin{equation}\label{eq4}
   \langle\hat{J}\rangle = \sum_n {c_n(t)}^2\langle n|\hat{J}| n\rangle + \sum_{n \neq m} c_n(t)c_m(t) e^{i(\gamma_n-\gamma_m)}\langle m|\hat{J}| n\rangle
  \end{equation}
   where $c_n(t)=|\langle n_\phi(t)|\psi(t)\rangle|$ describes the magnitude of the overlaps of the
time-dependent wave-function with the instantaneous eigenstates and $\gamma_n=arg(\langle
n_\phi(t)|\psi(t)\rangle)$ describe the phases acquired by the wave-function.

\begin{figure}[ttt]
\begin{center}
   \includegraphics[width=\columnwidth]{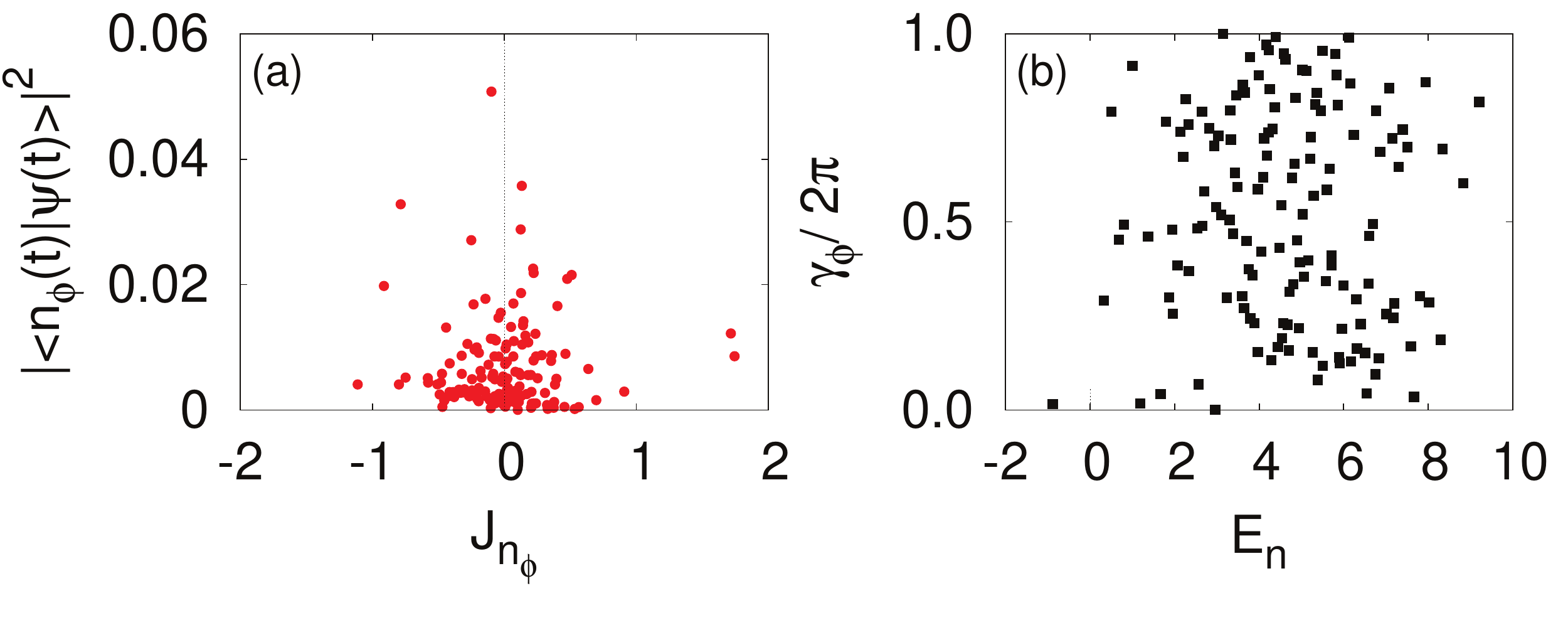}
   \caption {(Color online) (a) $|\langle n_\phi(t)|\psi(t)\rangle|$ for $L_{sites}=8$ and
$N_\uparrow=N_\downarrow=4, U=1.5, V=0.82$  and F=0.025 at $tF=2.51$, $|n_\phi(t)\rangle$ are
   the eigenstates of instantaneous Hamiltonian; (b) $\gamma_n=arg(\langle
n_\phi(t)|\psi(t)\rangle)$ for the same parameters.}
   \label{fig4}
   \vspace{-0.5cm}
\end{center}
\end{figure}

We plot in Fig.~\ref{fig4}(a), for $F=0.025$ and $L=8$, $c_n(t)$ as function of the current for each eigenstate at time $tF=2.51$. Observe that the probability amplitudes as function of current are approximately symmetrically distributed between left-going and right-going states, this in turn implies that the first term of Eq.~(\ref{eq4}), i.e. the diagonal expectation value of the current, becomes approximately equal to 0. Moreover, the phases, $
\gamma_n$, which
   are presented in Fig.~\ref{fig4}(b) are distributed uniformly between 0 and $2\pi$ therefore
leading to the dephasing of non-diagonal terms in Eq.~(\ref{eq4}),
    and finally the total current is approximately equal to zero. One should notice that for the case with $L=8$ the current is not completely equal to zero, but
    it acquires a small but finite value that fluctuates around zero, indicating the fact that the number of eigenstates that contribute is small due to finite size effects.
    These fluctuations are suppressed for larger systems as we show in the following sections.

  {\it Intermediate fields.} We next analyze the response to intermediate fields between the full dissipative case for $F=0.025$ and the full oscillating one for $F=0.4$. When the
   electric field strength is increased the reflection of the high probability state gradually approaches the largest eigenstate of the spectrum. This could be clearly recognized in Fig.~\ref{fig2} where
   the time, $tF$, for which the current changes its sign approaches $0.5$. At the same time the BO period, which is generally less than $F^{-1}$, gradually approaches $F^{-1}$. This is shown in the inset
   of Fig.~\ref{fig2} where we plot the frequency of BO as function of field strength. A similar
behavior was also reported in metallic spin-less systems subjected to an uniform electric
field\cite{PhysRevLett.105.186405}.
   Our investigation should also be relevant to that case. Similar to the
electric breakdown case, where a mapping to a quantum random
walk\cite{PhysRevLett.94.100602} on a semi-infinite chain was proposed, here the problem of BO
damping also could be mapped to
    a quantum random walk but on a chain with two edge states. However, as we will present in the
following, the actual long time response to an electric field depends strongly on the
probability transfer between subsequent states throughout the whole spectrum. It is therefore
necessary to design a random walk for which the probability transfer is also randomized but taken
from specific distributions, which could be chosen based on the level statistics of the
Hamiltonian\cite{poilblanc_poisson_1993}.

\begin{figure}[ttt]
\begin{center}
   \includegraphics[width=\columnwidth]{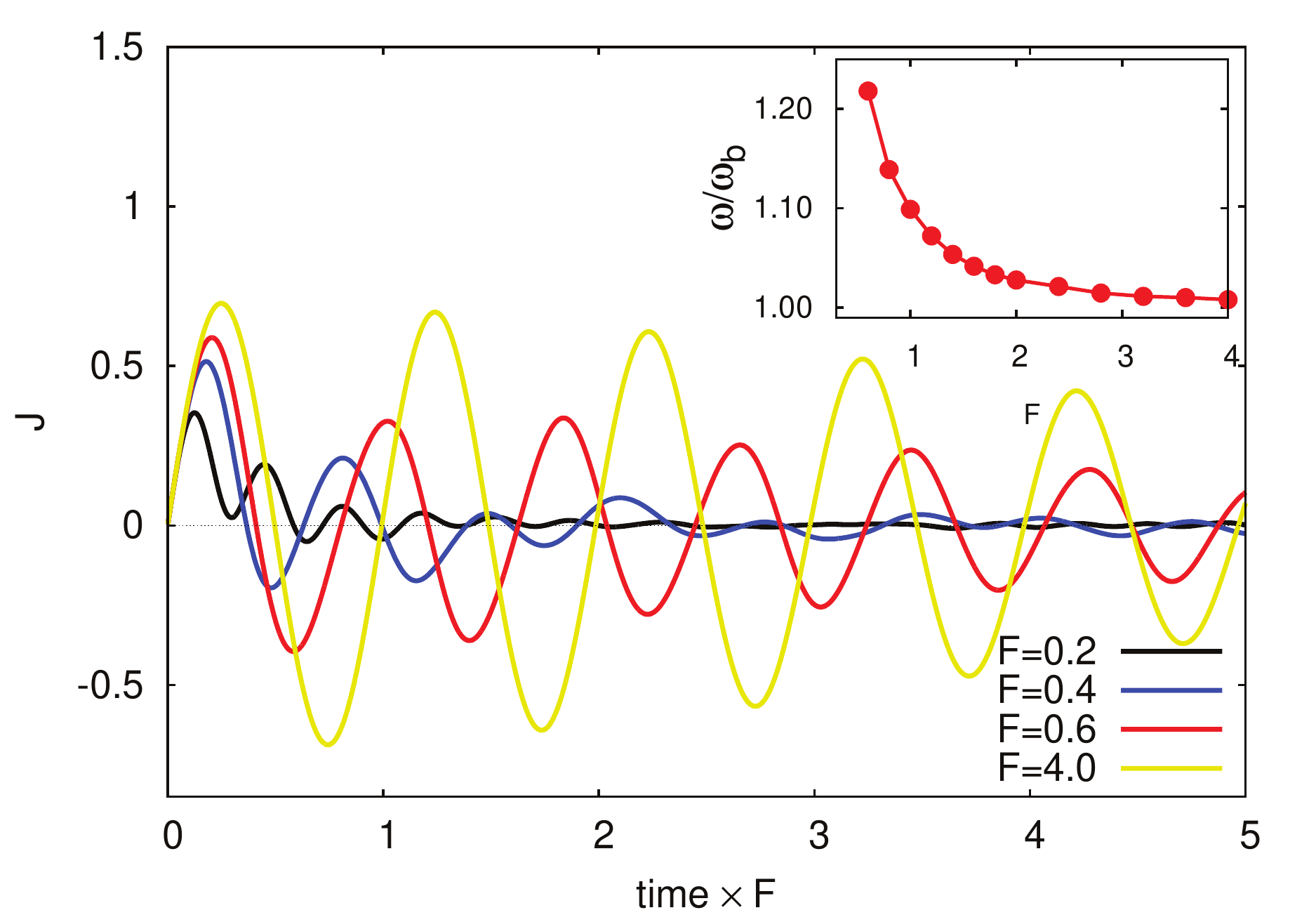}
   \caption {(Color online) Current as function of time for a dimerized ring with $L_{sites}=10$ and
$N_\uparrow=N_\downarrow=5, U=1.5, V=0.82, \eta=0.4$(see the definition of the hopping parameter
following Eq.~\ref{eq1}) and for different electric field strengths. The inset shows
the frequency of the Bloch oscillations for different electric fields and the same parameters of the main graph with $\omega_b=F$.}
   \label{fig5}
   \vspace{-0.5cm}
\end{center}
\end{figure}

{\it Dimerization}. In Fig.~\ref{fig5}, we show the current as function of time for a system with the same interactions as in the metallic case but with a dimerization parameter $\eta=0.4$. We call this state a correlated band insulator(BI). The general also arguments presented for the metallic case hold here, however there are also differences, which we explain in the following.
As expected, dimerization induces the opening of a charge gap ($\Delta_{charge}(\phi_{anti})=1.74$) and the electric field breakdown is postponed to larger fields. Additionally, a dissipative regime appears only at $F=0.2$. At this field strength the breakdown has already happened and the instantaneous ground state has a very small contribution to $|\psi(t)\rangle$. For larger fields, i.e. $F=0.4$, first the current starts to show irregular oscillations,  then at $F=0.6$ the current becomes oscillatory but with a frequency of the BO larger than $F$. Finally, at even larger fields, $F=4.0$, the current is oscillatory with $\omega=F$. This is achieved for much larger fields than the ones presented for the weakly interacting metal, as shown in the inset of Fig.~\ref{fig5}
The first notable difference between the metal and the correlated BI is that here BOs with smaller
frequencies survive for longer times. This is different from the metallic case where BOs with
smaller frequencies are strongly damped. Furthermore, one may expect that the dimerization may only
postpone the breakdown and the transition to the oscillatory behavior should not be affected as long
as the dimerization only affects the low energy part of the spectrum by opening up a ground-state
charge gap. However, the presence of long lasting BO with the period less than $F^{-1}$ implies the
presence of states with small $\chi_\varXi$ in the middle of the spectrum and which reflects a high
probability state back. Roughly speaking, these states could be at the edge of a cluster of
eigenstates, and are separated by a large gap from the next subsequent state and therefore play the
rule of an edge state. However, we emphasize that not only the gap but also the actual value of
$\chi_\varXi$ of each eigenstate are
important factor that affect the non-adiabatic behavior of the system.
In order to visualize again the overlap of the time-dependent wave-function with the whole spectrum,
we turn back to Fig.~\ref{fig3}(b), where the overlap with the instantaneous eigenstates of
$\hat{H}(t)$ is plotted as function of time for a smaller dimerized system with
$L=6$, $\eta=0.4$ and $F=2.0$. Again the smaller ring behaves the same as a larger system with
$L=10$ when subjected to strong fields. As is clear from the plot the recurrences of the
ground-state and the state with largest energy occur periodically at $F^{-1}$. A noticeable feature
of the propagation in the dimerized systems is the fact that the overlap of $|\psi(t)\rangle$ with
the instantaneous eigenstates is very nonlocal in the energy domain, meaning that the path of high
probability transition is broadened in comparison to the metallic system. Noticeably, the
wave-function starts to have finite overlap around the  first anti-crossing not only to the first
excited state but also with the second excited state. Therefore, a two level approximation (LZ-like)
is not appropriate for the ground-state breakdown.
The dimerization leads to a stronger insulator not only in the sense that it postpones the electric field breakdown, but it also largely affects the overlap with states located in the
middle of the spectrum. In short, while the breakdown and the appearance of the dissipative behavior mostly depends on the low energy part of the
spectrum, the transition from the dissipative to the oscillatory behavior largely depends on the clustering of eigenstates in the middle of the spectrum.
\subsection{Strongly interacting system}
\begin{figure}[ttt]
\begin{center}
   \includegraphics[width=\columnwidth]{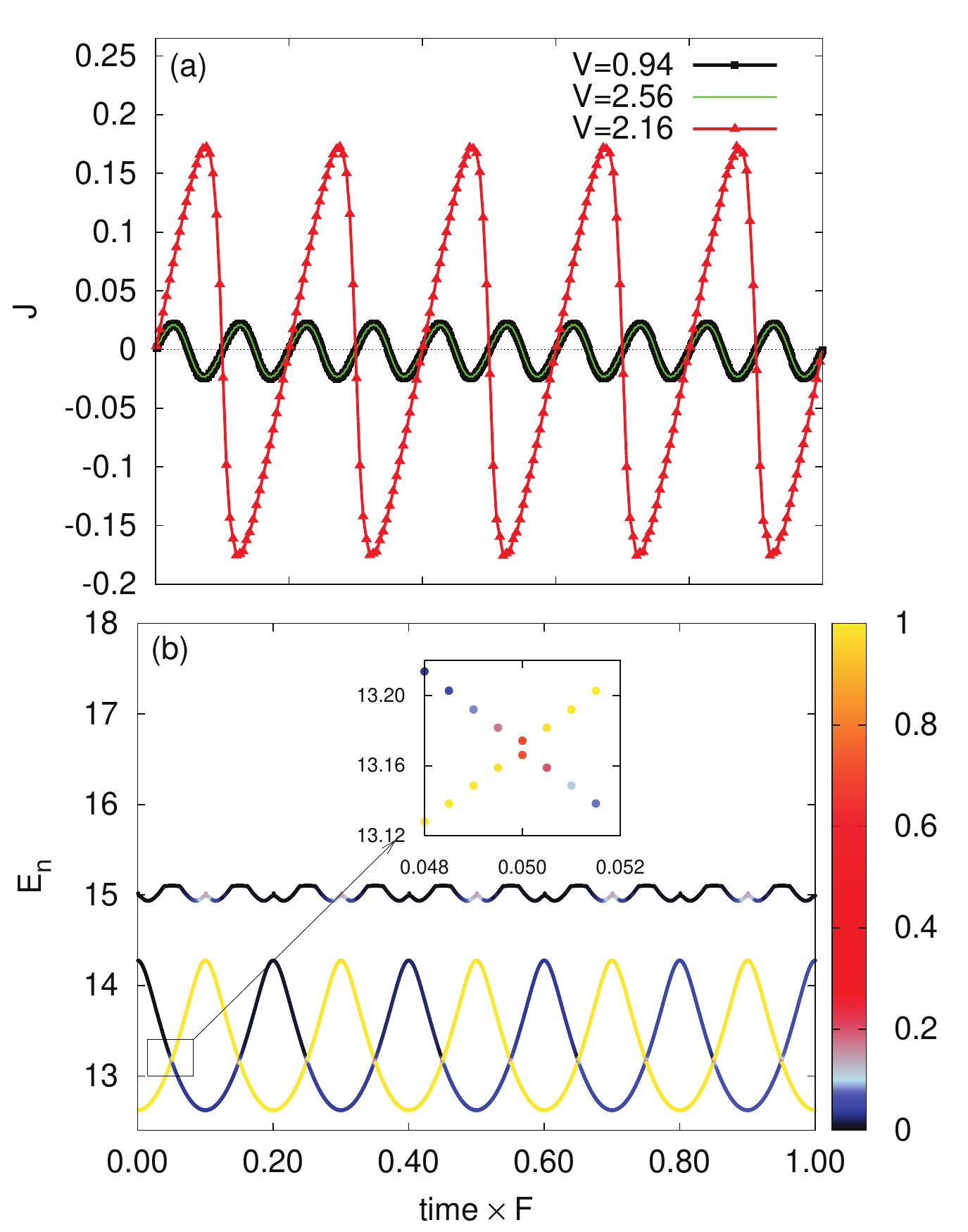}
   \caption {(Color online) (a) Current as function of time for very small field $F=0.002$ for
different interactions; (b) The energy dispersion of the first three exited
   states of instantaneous Hamiltonian together with the square of these states with
$|\psi(t)\rangle$ the as function of time for U=4.0,V=1.16.
   The inset of panel (b) shows a zoom-in into the into anti-crossing region. Colors represent the overlap of the time-dependent wave-function with the instantaneous eigenstates.}
   \label{fig6}
   \vspace{-0.5cm}
\end{center}
\end{figure}

For the cases with strong interactions, as stated before, we choose $U=4.0$ and three different nearest neighbor interactions, $V=0.94$ (SDW insulator), $V=2.56$ (CDW insulator) and $V=2.16$ (metallic case). For the insulating cases we choose the interaction parameters such that both cases acquire the same
ground state $\chi_\varXi(\phi_{anti})$ as seen in Fig.~\ref{fig2}.
We plot, in Fig.~\ref{fig6}(a), the current as function of time for a very small electric field, i.e. $F=0.002$, for a ring of size $L=10$. Both insulating systems appear to be in the adiabatic regime, where the current
shows an oscillatory behavior with a period equal to $F/L$. However, the metallic case shows oscillations with a doubled period, $2F/L$. The main reason for this comes from the fact that for the metallic case the
probability is transfered completely to the first excited state due to very large $\chi_\varXi(\phi_{anti})$, i.e. it cannot be considered in the adiabatic regime even at these small fields. This is illustrated in
Fig.~\ref{fig6}(b), where the energies of the first three states of the $\hat{H}(t)$ are shown as function of time (and implicitly as a function of flux), together with the overlap of $|\psi(t)\rangle$ to these three states. As is obvious, because of the very large  $\chi_\varXi$,
there is a very large overlap to the first excited state after the first anti-crossing, however the field is very small such that it cannot overcome the gap between the first excited
 and second excited state. $|\psi(t)\rangle$ only has an extremely small overlap with second excited state, which leads to the fact that the probability
 is reflected back to the ground state and one ends up with current oscillations with a period twice of the adiabatic expectation. The breakdown field is now achieved when the gap between the first and second excited
 states is overcome.
 We next describe the response of strongly interacting systems to larger fields. In Fig.~\ref{fig6} we present the current as function of time for different field strengths and for the three interaction choices introduced previously. For $F=0.1$ all the
 cases shows a dissipative behavior, however the insulating ones show small peaks in the current before it arrives at the quasi-stationary zero-current state. The period of these peaks is approximately equal to $F/L$, which therefore implies that the overlap of $|\psi(t)\rangle$ with the instantaneous ground state does not vanish quickly and manifest itself as small peaks in the current. This is not the case for $V=2.16$ where the overlap with the ground state
 is lost immediately at the anti-crossing (see the inset of Fig.~\ref{fig6} for $F=0.2$) and the current behaves smoothly from the beginning of
 the evolution. For stronger fields, $F=0.2$, the change of the current is large, such that the
current fluctuations due to the finite overlap with the ground-state disappear. In the inset of
Fig.~\ref{fig6} we show the square of the overlap of
$|\psi(t)\rangle$ with the ground-state of the instantaneous Hamiltonian. It is
 clear that for the two insulating cases for which we set $\chi_\varXi(\phi_{anti})$ to be equal,
the decay of the ground-state is identical. Furthermore, in the dissipative cases for $F=0.1$ and
$F=0.2$ both insulating cases behave almost in
 the same way even for larger times even though the interaction strengths are very different and one describes an SDW insulator while the other one an CDW insulator with different excitation. This means that by
 setting $\chi_\varXi(\phi_{anti})$ the same, not only the ground-state decay is identical but also
the tunneling to the lower part of spectrum the behaves very similarly. When the field is increased
to $F=0.8$, the SDW
 insulator with $V=0.94$ starts to show Bloch-like oscillations with large amplitude. On the other hand the metallic and CDW cases  are still in the dissipative regime with a vanishingly small long-time current. For even
 larger fields, $F\simeq1.6$ (not shown here), all three cases show oscillations with large amplitude but which are irregular. It is only when the strength of the electric field is very large, $F=10.0$, that all the cases
 show regular BO oscillations as shown in Fig.~\ref{fig6}.
\begin{figure}[ttt]
\begin{center}
   \includegraphics[width=\columnwidth]{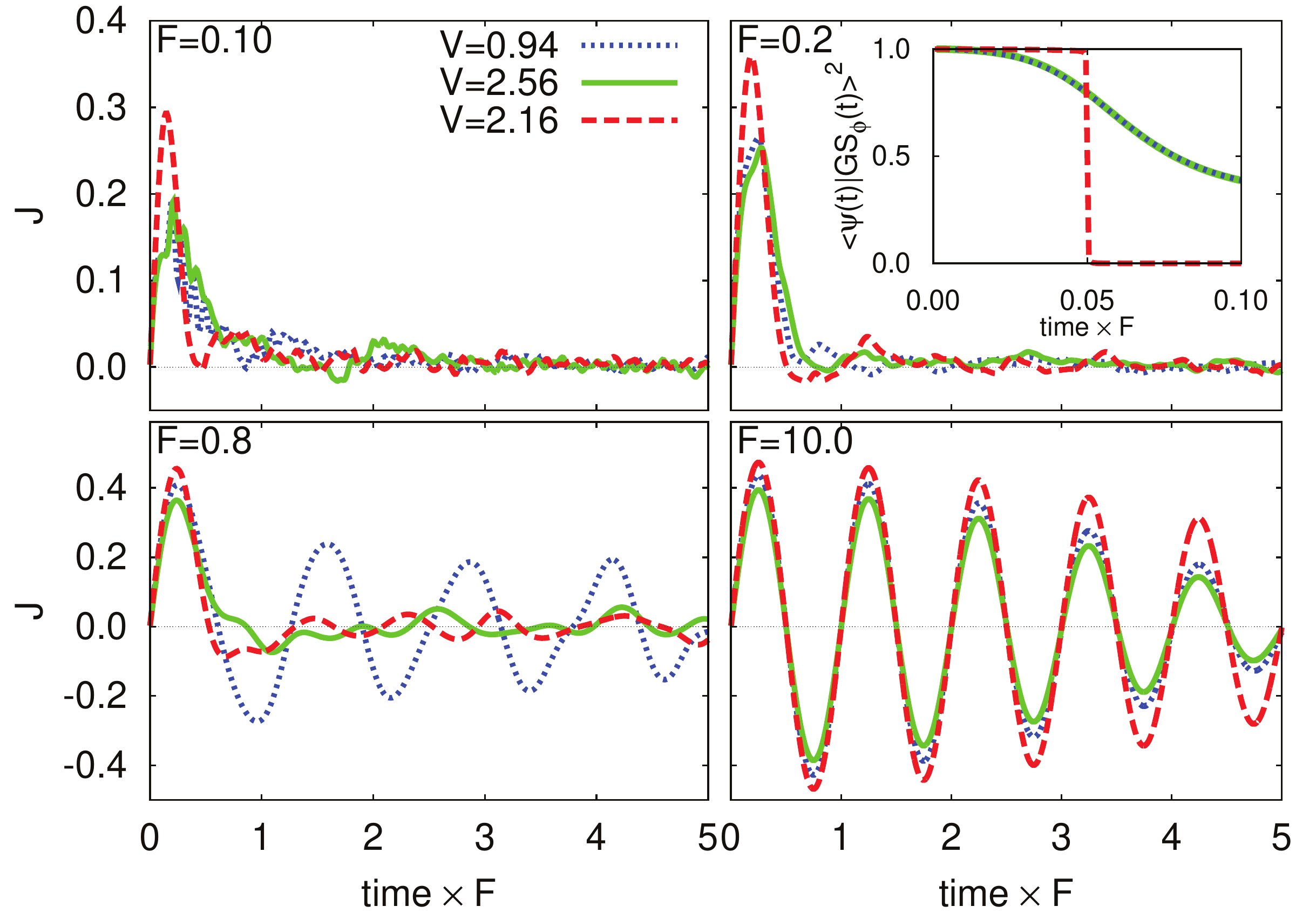}
     \caption {(Color online) Current as function of time for different interactions and different
field strength. The inset shows the square of the overlap of
     $|\psi(t)\rangle$ with the instantaneous ground state of $H(t)$ for different interactions and $F=2.0$.}
   \label{fig7}
   \vspace{-0.5cm}
\end{center}
\end{figure}

\begin{figure}[ttt]
\begin{center}
   \includegraphics[width=\columnwidth]{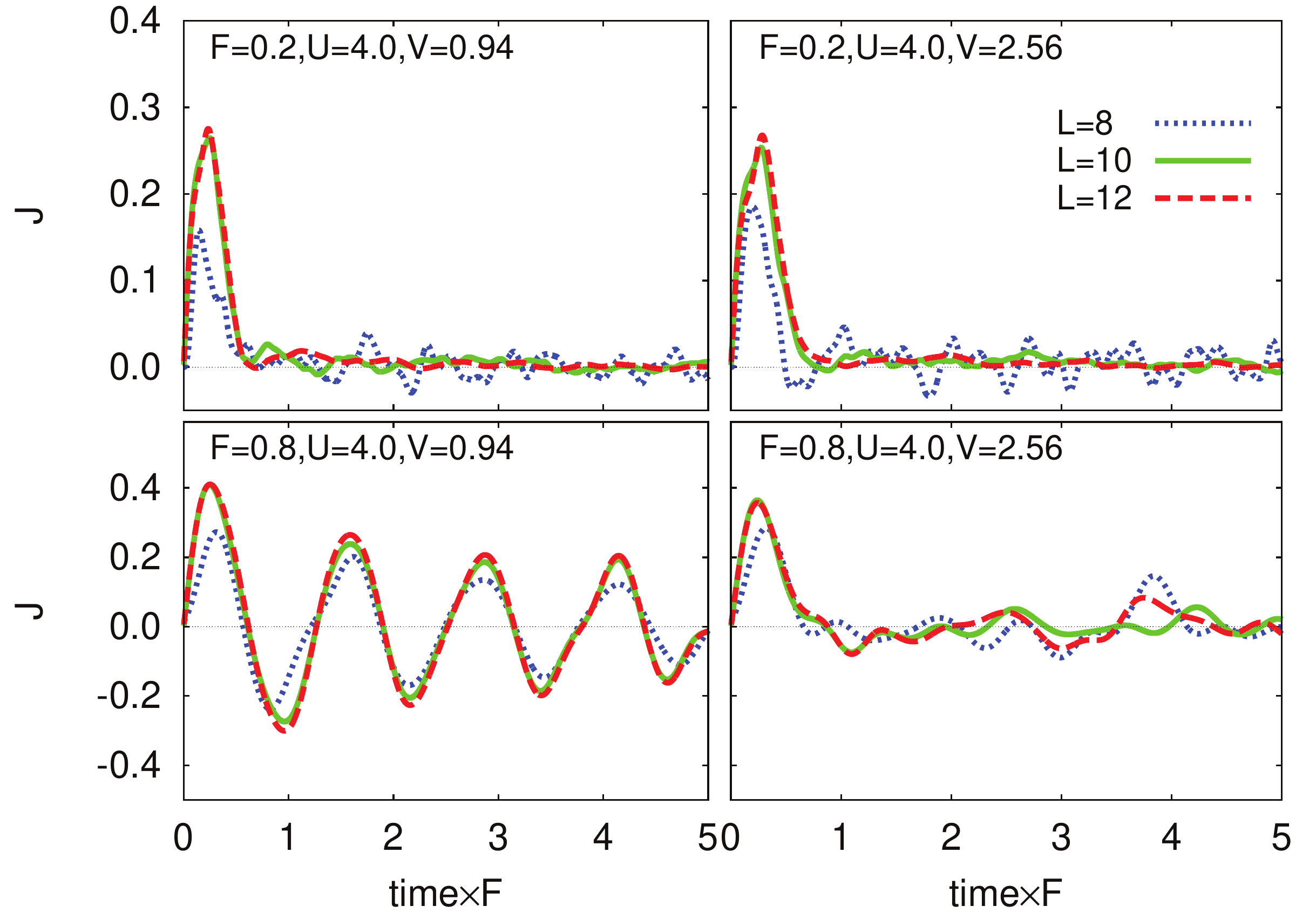}
     \caption {(Color online) Current as function of time for different interactions and different
field strengths and different sizes.}
   \label{fig8}
   \vspace{-0.5cm}
\end{center}
\end{figure}

 {\it Finite size effect.} To see the effect of the size of the system on the transition from a
dissipative to an oscillatory pattern, we plot in Fig.~\ref{fig8} the current
  as function of time for different sizes for $V=0.94$ (SDW insulator) and $V=2.56$ (CDW insulator). We observe that for all cases the fluctuations
   of the current in the dissipative regime ($F=0.2$) are suppressed for larger sizes. This is due to the fact that $|\psi(t)\rangle$ acquires overlap with a much larger
   number of states when the size is increased. This implies that a more efficient dephasing of the current is achieved (see the discussion
    following Eq.~\ref{eq4}). However, in the strong-field regime, once the transition to oscillatory behavior occurs, the size effect is negligible,
showing that the sizes of the gaps in the middle of the spectrum do not depend strongly on the size, at least not for the strong interactions considered here.

\begin{figure}[ttt]
\begin{center}
   \includegraphics[width=\columnwidth]{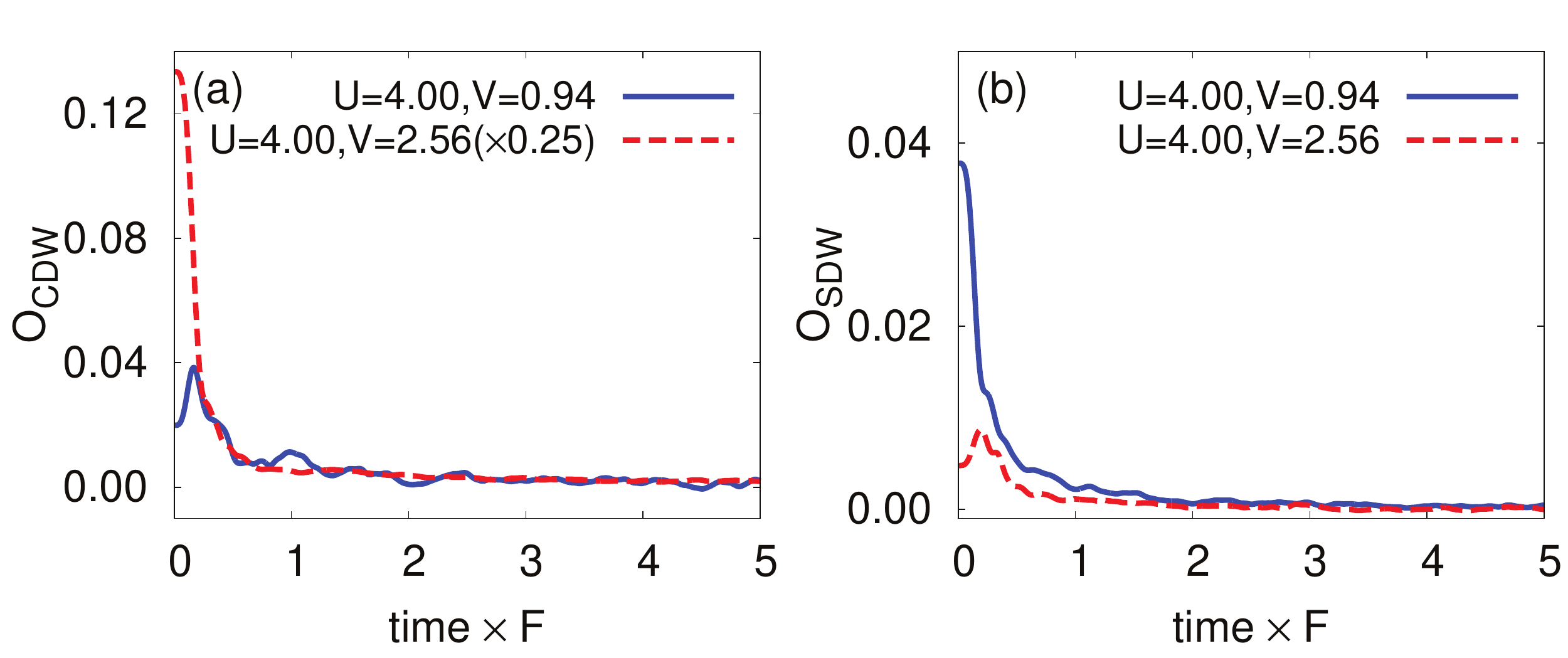}
     \caption {(Color online) (a) CDW order parameter as function of time for a system with L=12 at
half filling derived with $F=0.2$; (b) SDW order parameter
       as function of time and the same parameters as in plot(a).}
   \label{fig9}
   \vspace{-0.5cm}
\end{center}
\end{figure}

{\it Order parameters.} In Fig.~\ref{fig9}, we show the SDW and CDW order parameters as function of
time for the two insulating cases. As is clear
    from Fig.~\ref{fig9}(b) for the SDW ordered system, O$_{SDW}$ only drops gradually as
function of time, however at the same time
    O$_{CDW}$ is enhanced at the beginning of the evolution. This further implies the presence of
a CDW state near the bottom of the spectrum\cite{lu_enhanced_2012}. Finally at longer times
both order parameters dissipates during the evolution arriving at a quasi-stationary
     state with almost vanishing value for larger times. The CDW ordered system shows a similar
behavior but with reversed $O_{CDW}$ and $O_{SDW}$ contributions (see Fig.~\ref{fig9}(b)). Therefore
, the transient regime shows that since the two order parameters are in competition, the mechanism
of destroying
the dominant order is the proliferation of the competing one.
\section{Conclusions}
 In conclusion we investigated the nonlinear response of a closed interacting fermionic system
as modeled by an extended Hubbard model. Weakly
 interacting metallic system at the boundary of SDW-CDW, shows a dissipative behavior for low fields. The main reason for this is the fact that
  $|\psi(t)\rangle$ acquires overlap with large number of left going and right going states. This
in turn implies that the quasi-stationary state acquires zero current. Bloch oscillations start
smoothly with a frequency larger than
  $F$. The main reason for this is the fact that the reflection happens at the
   lower part of the spectrum, thus effectively decreasing the bandwidth. Upon increasing the field
strength the probability transfer at each anti-crossing is more efficient. This leads to
  a more regular recurrences of the ground state and the topmost excited state with period of $F^{-1}$,
which can be seen from the oscillations in the current. Upon dimerization of the metallic system,
the formation of the dissipative regime is postponed to larger fields due to the formation of a charge
  gap. However, the main difference between the dimer case and the metallic system resides in the
fact that, first, it shows irregular current oscillations
  before they turn into regular BO and second, the BO with larger frequencies survive for large
times, in analogy with the metallic case subjected to strong field.
  This implies the existence of states at the middle of the spectrum with low $\chi_\varXi(\phi)$
(or roughly speaking the formation of large mid gaps in the relevant excitations) that play the role
of a band-edge state and reflect back the overlap probability at the middle part of the spectrum
even for large electric fields. Finally, the dimerized system also shows regular BO with period
equal to $F^{-1}$ for large enough electric fields. The value for which the dimerized system shows
regular BO are much larger than those found for the metallic system even though the interactions are
identical.

For stronger interacting systems when the interactions are chosen such that the ground-state
$\chi_\varXi(\phi_{anti})$ is the same for both cases, then the ground state decay for both CDW
insulator and SDW insulator behaves exactly the same. This similarity of the ground-state decay
manifests itself even for larger times and for both low and high field dissipative regimes.
However significant differences arise
  between the two cases for large electric fields. While SDW shows oscillatory behavior with large
magnitude the CDW insulator and strong interacting metallic system
  only shows irregularities with small oscillations. Different from the weakly interacting
metallic system and the dimer case, in the strongly interacting regime these
  irregularities are extended to intermediate fields and only for very large fields, $F=10.0$,
regular BO with a period of $F^{-1}$ are observed. This effect
   appears to be little affected by size, since the SDW and CDW
insulators, for L=10 and L=12, show the same qualitatively and even quantitatively behavior. This
implies that the reorganization of the spectrum is affected much more by the interaction than by
the finite size induced discreteness.
\begin{acknowledgements}
 This work was supported by the Flemish Science Foundation (FWO-Vl)and Methusalem program of the Flemish government. One of us (LC) is a postdoctoral fellow of the FWO-Vl.
\end{acknowledgements}
\bibliographystyle{apsrev4-1}
%{apsrev4-1}
%\bibliography{myref1}

\begin{thebibliography}{46}%
\makeatletter
\providecommand \@ifxundefined [1]{%
 \@ifx{#1\undefined}
}%
\providecommand \@ifnum [1]{%
 \ifnum #1\expandafter \@firstoftwo
 \else \expandafter \@secondoftwo
 \fi
}%
\providecommand \@ifx [1]{%
 \ifx #1\expandafter \@firstoftwo
 \else \expandafter \@secondoftwo
 \fi
}%
\providecommand \natexlab [1]{#1}%
\providecommand \enquote  [1]{``#1''}%
\providecommand \bibnamefont  [1]{#1}%
\providecommand \bibfnamefont [1]{#1}%
\providecommand \citenamefont [1]{#1}%
\providecommand \href@noop [0]{\@secondoftwo}%
\providecommand \href [0]{\begingroup \@sanitize@url \@href}%
\providecommand \@href[1]{\@@startlink{#1}\@@href}%
\providecommand \@@href[1]{\endgroup#1\@@endlink}%
\providecommand \@sanitize@url [0]{\catcode `\\12\catcode `\$12\catcode
  `\&12\catcode `\#12\catcode `\^12\catcode `\_12\catcode `\%12\relax}%
\providecommand \@@startlink[1]{}%
\providecommand \@@endlink[0]{}%
\providecommand \url  [0]{\begingroup\@sanitize@url \@url }%
\providecommand \@url [1]{\endgroup\@href {#1}{\urlprefix }}%
\providecommand \urlprefix  [0]{URL }%
\providecommand \Eprint [0]{\href }%
\providecommand \doibase [0]{http://dx.doi.org/}%
\providecommand \selectlanguage [0]{\@gobble}%
\providecommand \bibinfo  [0]{\@secondoftwo}%
\providecommand \bibfield  [0]{\@secondoftwo}%
\providecommand \translation [1]{[#1]}%
\providecommand \BibitemOpen [0]{}%
\providecommand \bibitemStop [0]{}%
\providecommand \bibitemNoStop [0]{.\EOS\space}%
\providecommand \EOS [0]{\spacefactor3000\relax}%
\providecommand \BibitemShut  [1]{\csname bibitem#1\endcsname}%
\let\auto@bib@innerbib\@empty
%</preamble>
\bibitem [{\citenamefont {Polkovnikov}\ \emph {et~al.}(2011)\citenamefont
  {Polkovnikov}, \citenamefont {Sengupta}, \citenamefont {Silva},\ and\
  \citenamefont {Vengalattore}}]{RevModPhys.83.863}%
  \BibitemOpen
  \bibfield  {author} {\bibinfo {author} {\bibfnamefont {A.}~\bibnamefont
  {Polkovnikov}}, \bibinfo {author} {\bibfnamefont {K.}~\bibnamefont
  {Sengupta}}, \bibinfo {author} {\bibfnamefont {A.}~\bibnamefont {Silva}}, \
  and\ \bibinfo {author} {\bibfnamefont {M.}~\bibnamefont {Vengalattore}},\
  }\href {\doibase 10.1103/RevModPhys.83.863} {\bibfield  {journal} {\bibinfo
  {journal} {Rev. Mod. Phys.}\ }\textbf {\bibinfo {volume} {83}},\ \bibinfo
  {pages} {863} (\bibinfo {year} {2011})}\BibitemShut {NoStop}%
\bibitem [{\citenamefont {Wei\ss{}e}\ \emph {et~al.}(2006)\citenamefont
  {Wei\ss{}e}, \citenamefont {Wellein}, \citenamefont {Alvermann},\ and\
  \citenamefont {Fehske}}]{RevModPhys.78.275}%
  \BibitemOpen
  \bibfield  {author} {\bibinfo {author} {\bibfnamefont {A.}~\bibnamefont
  {Wei\ss{}e}}, \bibinfo {author} {\bibfnamefont {G.}~\bibnamefont {Wellein}},
  \bibinfo {author} {\bibfnamefont {A.}~\bibnamefont {Alvermann}}, \ and\
  \bibinfo {author} {\bibfnamefont {H.}~\bibnamefont {Fehske}},\ }\href
  {\doibase 10.1103/RevModPhys.78.275} {\bibfield  {journal} {\bibinfo
  {journal} {Rev. Mod. Phys.}\ }\textbf {\bibinfo {volume} {78}},\ \bibinfo
  {pages} {275} (\bibinfo {year} {2006})}\BibitemShut {NoStop}%
\bibitem [{\citenamefont {Hochbruck}\ and\ \citenamefont
  {Lubich}(1997)}]{hochbruck_krylov_1997}%
  \BibitemOpen
  \bibfield  {author} {\bibinfo {author} {\bibfnamefont {M.}~\bibnamefont
  {Hochbruck}}\ and\ \bibinfo {author} {\bibfnamefont {C.}~\bibnamefont
  {Lubich}},\ }\href {\doibase 10.1137/S0036142995280572} {\bibfield  {journal}
  {\bibinfo  {journal} {{SIAM} Journal on Numerical Analysis}\ }\textbf
  {\bibinfo {volume} {34}},\ \bibinfo {pages} {1911} (\bibinfo {year}
  {1997})}\BibitemShut {NoStop}%
\bibitem [{\citenamefont {Verstraete}\ \emph {et~al.}(2004)\citenamefont
  {Verstraete}, \citenamefont {Garcia-Ripoll},\ and\ \citenamefont
  {Cirac}}]{PhysRevLett.93.207204}%
  \BibitemOpen
  \bibfield  {author} {\bibinfo {author} {\bibfnamefont {F.}~\bibnamefont
  {Verstraete}}, \bibinfo {author} {\bibfnamefont {J.~J.}\ \bibnamefont
  {Garcia-Ripoll}}, \ and\ \bibinfo {author} {\bibfnamefont {J.~I.}\ \bibnamefont
  {Cirac}},\ }\href {\doibase 10.1103/PhysRevLett.93.207204} {\bibfield
  {journal} {\bibinfo  {journal} {Phys. Rev. Lett.}\ }\textbf {\bibinfo
  {volume} {93}},\ \bibinfo {pages} {207204} (\bibinfo {year}
  {2004})}\BibitemShut {NoStop}%
\bibitem [{\citenamefont {Freericks}(2008)}]{PhysRevB.77.075109}%
  \BibitemOpen
  \bibfield  {author} {\bibinfo {author} {\bibfnamefont {J.~K.}\ \bibnamefont
  {Freericks}},\ }\href {\doibase 10.1103/PhysRevB.77.075109} {\bibfield
  {journal} {\bibinfo  {journal} {Phys. Rev. B}\ }\textbf {\bibinfo {volume}
  {77}},\ \bibinfo {pages} {075109} (\bibinfo {year} {2008})}\BibitemShut
  {NoStop}%
\bibitem [{\citenamefont {Eckstein}\ and\ \citenamefont
  {Werner}(2013)}]{PhysRevB.88.075135}%
  \BibitemOpen
  \bibfield  {author} {\bibinfo {author} {\bibfnamefont {M.}~\bibnamefont
  {Eckstein}}\ and\ \bibinfo {author} {\bibfnamefont {P.}~\bibnamefont
  {Werner}},\ }\href {\doibase 10.1103/PhysRevB.88.075135} {\bibfield
  {journal} {\bibinfo  {journal} {Phys. Rev. B}\ }\textbf {\bibinfo {volume}
  {88}},\ \bibinfo {pages} {075135} (\bibinfo {year} {2013})}\BibitemShut
  {NoStop}%
\bibitem [{\citenamefont {Sawano}\ \emph {et~al.}(2005)\citenamefont {Sawano},
  \citenamefont {Terasaki}, \citenamefont {Mori}, \citenamefont {Mori},
  \citenamefont {Watanabe}, \citenamefont {Ikeda}, \citenamefont {Nogami},\
  and\ \citenamefont {Noda}}]{sawano_organic_2005}%
  \BibitemOpen
  \bibfield  {author} {\bibinfo {author} {\bibfnamefont {F.}~\bibnamefont
  {Sawano}}, \bibinfo {author} {\bibfnamefont {I.}~\bibnamefont {Terasaki}},
  \bibinfo {author} {\bibfnamefont {H.}~\bibnamefont {Mori}}, \bibinfo {author}
  {\bibfnamefont {T.}~\bibnamefont {Mori}}, \bibinfo {author} {\bibfnamefont
  {M.}~\bibnamefont {Watanabe}}, \bibinfo {author} {\bibfnamefont
  {N.}~\bibnamefont {Ikeda}}, \bibinfo {author} {\bibfnamefont
  {Y.}~\bibnamefont {Nogami}}, \ and\ \bibinfo {author} {\bibfnamefont
  {Y.}~\bibnamefont {Noda}},\ }\href {\doibase 10.1038/nature04087} {\bibfield
  {journal} {\bibinfo  {journal} {Nature}\ }\textbf {\bibinfo {volume} {437}},\
  \bibinfo {pages} {522} (\bibinfo {year} {2005})}\BibitemShut {NoStop}%
\bibitem [{\citenamefont {Taguchi}\ \emph {et~al.}(2000)\citenamefont
  {Taguchi}, \citenamefont {Matsumoto},\ and\ \citenamefont
  {Tokura}}]{taguchi_dielectric_2000}%
  \BibitemOpen
  \bibfield  {author} {\bibinfo {author} {\bibfnamefont {Y.}~\bibnamefont
  {Taguchi}}, \bibinfo {author} {\bibfnamefont {T.}~\bibnamefont {Matsumoto}},
  \ and\ \bibinfo {author} {\bibfnamefont {Y.}~\bibnamefont {Tokura}},\ }\href
  {\doibase 10.1103/PhysRevB.62.7015} {\bibfield  {journal} {\bibinfo
  {journal} {Phys. Rev. B}\ }\textbf {\bibinfo {volume} {62}},\ \bibinfo
  {pages} {7015} (\bibinfo {year} {2000})}\BibitemShut {NoStop}%
\bibitem [{\citenamefont {Iwai}\ \emph {et~al.}(2003)\citenamefont {Iwai},
  \citenamefont {Ono}, \citenamefont {Maeda}, \citenamefont {Matsuzaki},
  \citenamefont {Kishida}, \citenamefont {Okamoto},\ and\ \citenamefont
  {Tokura}}]{iwai_ultrafast_2003}%
  \BibitemOpen
  \bibfield  {author} {\bibinfo {author} {\bibfnamefont {S.}~\bibnamefont
  {Iwai}}, \bibinfo {author} {\bibfnamefont {M.}~\bibnamefont {Ono}}, \bibinfo
  {author} {\bibfnamefont {A.}~\bibnamefont {Maeda}}, \bibinfo {author}
  {\bibfnamefont {H.}~\bibnamefont {Matsuzaki}}, \bibinfo {author}
  {\bibfnamefont {H.}~\bibnamefont {Kishida}}, \bibinfo {author} {\bibfnamefont
  {H.}~\bibnamefont {Okamoto}}, \ and\ \bibinfo {author} {\bibfnamefont
  {Y.}~\bibnamefont {Tokura}},\ }\href {\doibase 10.1103/PhysRevLett.91.057401}
  {\bibfield  {journal} {\bibinfo  {journal} {Phys. Rev. Lett.}\ }\textbf
  {\bibinfo {volume} {91}},\ \bibinfo {pages} {057401} (\bibinfo {year}
  {2003})}\BibitemShut {NoStop}%
\bibitem [{\citenamefont {Okamoto}\ \emph {et~al.}(2007)\citenamefont
  {Okamoto}, \citenamefont {Matsuzaki}, \citenamefont {Wakabayashi},
  \citenamefont {Takahashi},\ and\ \citenamefont
  {Hasegawa}}]{okamoto_photoinduced_2007}%
  \BibitemOpen
  \bibfield  {author} {\bibinfo {author} {\bibfnamefont {H.}~\bibnamefont
  {Okamoto}}, \bibinfo {author} {\bibfnamefont {H.}~\bibnamefont {Matsuzaki}},
  \bibinfo {author} {\bibfnamefont {T.}~\bibnamefont {Wakabayashi}}, \bibinfo
  {author} {\bibfnamefont {Y.}~\bibnamefont {Takahashi}}, \ and\ \bibinfo
  {author} {\bibfnamefont {T.}~\bibnamefont {Hasegawa}},\ }\href {\doibase
  10.1103/PhysRevLett.98.037401} {\bibfield  {journal} {\bibinfo  {journal}
  {Phys. Rev. Lett.}\ }\textbf {\bibinfo {volume} {98}},\ \bibinfo {pages}
  {037401} (\bibinfo {year} {2007})}\BibitemShut {NoStop}%
\bibitem [{\citenamefont {Greiner}\ \emph {et~al.}(2002)\citenamefont
  {Greiner}, \citenamefont {Mandel}, \citenamefont {Hansch},\ and\
  \citenamefont {Bloch}}]{nature_optical}%
  \BibitemOpen
  \bibfield  {author} {\bibinfo {author} {\bibfnamefont {M.}~\bibnamefont
  {Greiner}}, \bibinfo {author} {\bibfnamefont {O.}~\bibnamefont {Mandel}},
  \bibinfo {author} {\bibfnamefont {T.~W.}\ \bibnamefont {Hansch}}, \ and\
  \bibinfo {author} {\bibfnamefont {I.}~\bibnamefont {Bloch}},\ }\href@noop {}
  {\bibfield  {journal} {\bibinfo  {journal} {Nature}\ }\textbf {\bibinfo
  {volume} {419}},\ \bibinfo {pages} {51} (\bibinfo {year} {2002})}\BibitemShut
  {NoStop}%
\bibitem [{\citenamefont {K\"ohl}\ \emph {et~al.}(2005)\citenamefont {K\"ohl},
  \citenamefont {Moritz}, \citenamefont {St\"oferle}, \citenamefont
  {G\"unter},\ and\ \citenamefont {Esslinger}}]{PhysRevLett.94.080403}%
  \BibitemOpen
  \bibfield  {author} {\bibinfo {author} {\bibfnamefont {M.}~\bibnamefont
  {K\"ohl}}, \bibinfo {author} {\bibfnamefont {H.}~\bibnamefont {Moritz}},
  \bibinfo {author} {\bibfnamefont {T.}~\bibnamefont {St\"oferle}}, \bibinfo
  {author} {\bibfnamefont {K.}~\bibnamefont {G\"unter}}, \ and\ \bibinfo
  {author} {\bibfnamefont {T.}~\bibnamefont {Esslinger}},\ }\href {\doibase
  10.1103/PhysRevLett.94.080403} {\bibfield  {journal} {\bibinfo  {journal}
  {Phys. Rev. Lett.}\ }\textbf {\bibinfo {volume} {94}},\ \bibinfo {pages}
  {080403} (\bibinfo {year} {2005})}\BibitemShut {NoStop}%
\bibitem [{\citenamefont {Bloch}(2005)}]{bloch_ultracold_2005}%
  \BibitemOpen
  \bibfield  {author} {\bibinfo {author} {\bibfnamefont {I.}~\bibnamefont
  {Bloch}},\ }\href {\doibase 10.1038/nphys138} {\bibfield  {journal} {\bibinfo
   {journal} {Nat Phys}\ }\textbf {\bibinfo {volume} {1}},\ \bibinfo {pages}
  {23} (\bibinfo {year} {2005})}\BibitemShut {NoStop}%
\bibitem [{\citenamefont {Esslinger}(2010)}]{esslinger_fermi-hubbard_2010}%
  \BibitemOpen
  \bibfield  {author} {\bibinfo {author} {\bibfnamefont {T.}~\bibnamefont
  {Esslinger}},\ }\href {\doibase 10.1146/annurev-conmatphys-070909-104059}
  {\bibfield  {journal} {\bibinfo  {journal} {Annual Review of Condensed Matter
  Physics}\ }\textbf {\bibinfo {volume} {1}},\ \bibinfo {pages} {129} (\bibinfo
  {year} {2010})}\BibitemShut {NoStop}%
\bibitem [{\citenamefont {Eckstein}\ \emph
  {et~al.}(2010{\natexlab{a}})\citenamefont {Eckstein}, \citenamefont
  {Kollar},\ and\ \citenamefont {Werner}}]{eckstein_interaction_2010}%
  \BibitemOpen
  \bibfield  {author} {\bibinfo {author} {\bibfnamefont {M.}~\bibnamefont
  {Eckstein}}, \bibinfo {author} {\bibfnamefont {M.}~\bibnamefont {Kollar}}, \
  and\ \bibinfo {author} {\bibfnamefont {P.}~\bibnamefont {Werner}},\ }\href
  {\doibase 10.1103/PhysRevB.81.115131} {\bibfield  {journal} {\bibinfo
  {journal} {Phys. Rev. B}\ }\textbf {\bibinfo {volume} {81}},\ \bibinfo
  {pages} {115131} (\bibinfo {year} {2010}{\natexlab{a}})}\BibitemShut
  {NoStop}%
\bibitem [{\citenamefont {Eckstein}\ \emph {et~al.}(2009)\citenamefont
  {Eckstein}, \citenamefont {Hackl}, \citenamefont {Kehrein}, \citenamefont
  {Kollar}, \citenamefont {Moeckel}, \citenamefont {Werner},\ and\
  \citenamefont {Wolf}}]{eckstein_new_2009}%
  \BibitemOpen
  \bibfield  {author} {\bibinfo {author} {\bibfnamefont {M.}~\bibnamefont
  {Eckstein}}, \bibinfo {author} {\bibfnamefont {A.}~\bibnamefont {Hackl}},
  \bibinfo {author} {\bibfnamefont {S.}~\bibnamefont {Kehrein}}, \bibinfo
  {author} {\bibfnamefont {M.}~\bibnamefont {Kollar}}, \bibinfo {author}
  {\bibfnamefont {M.}~\bibnamefont {Moeckel}}, \bibinfo {author} {\bibfnamefont
  {P.}~\bibnamefont {Werner}}, \ and\ \bibinfo {author} {\bibfnamefont {F.~A.}\
  \bibnamefont {Wolf}},\ }\href {\doibase 10.1140/epjst/e2010-01219-x}
  {\bibfield  {journal} {\bibinfo  {journal} {Eur. Phys. J. Spec. Top.}\
  }\textbf {\bibinfo {volume} {180}},\ \bibinfo {pages} {217} (\bibinfo {year}
  {2009})}\BibitemShut {NoStop}%
\bibitem [{\citenamefont {Kollath}\ \emph {et~al.}(2007)\citenamefont
  {Kollath}, \citenamefont {Lauchli},\ and\ \citenamefont
  {Altman}}]{PhysRevLett.98.180601}%
  \BibitemOpen
  \bibfield  {author} {\bibinfo {author} {\bibfnamefont {C.}~\bibnamefont
  {Kollath}}, \bibinfo {author} {\bibfnamefont {A.~M.}\ \bibnamefont
  {Lauchli}}, \ and\ \bibinfo {author} {\bibfnamefont {E.}~\bibnamefont
  {Altman}},\ }\href {\doibase 10.1103/PhysRevLett.98.180601} {\bibfield
  {journal} {\bibinfo  {journal} {Phys. Rev. Lett.}\ }\textbf {\bibinfo
  {volume} {98}},\ \bibinfo {pages} {180601} (\bibinfo {year}
  {2007})}\BibitemShut {NoStop}%
\bibitem [{\citenamefont {Genway}\ \emph {et~al.}(2012)\citenamefont {Genway},
  \citenamefont {Ho},\ and\ \citenamefont {Lee}}]{genway_thermalization_2012}%
  \BibitemOpen
  \bibfield  {author} {\bibinfo {author} {\bibfnamefont {S.}~\bibnamefont
  {Genway}}, \bibinfo {author} {\bibfnamefont {A.~F.}\ \bibnamefont {Ho}}, \
  and\ \bibinfo {author} {\bibfnamefont {D.~K.~K.}\ \bibnamefont {Lee}},\
  }\href {\doibase 10.1103/PhysRevA.86.023609} {\bibfield  {journal} {\bibinfo
  {journal} {Phys. Rev. A}\ }\textbf {\bibinfo {volume} {86}},\ \bibinfo
  {pages} {023609} (\bibinfo {year} {2012})}\BibitemShut {NoStop}%
\bibitem [{\citenamefont {Rigol}\ and\ \citenamefont
  {Santos}(2010)}]{PhysRevA.82.011604}%
  \BibitemOpen
  \bibfield  {author} {\bibinfo {author} {\bibfnamefont {M.}~\bibnamefont
  {Rigol}}\ and\ \bibinfo {author} {\bibfnamefont {L.~F.}\ \bibnamefont
  {Santos}},\ }\href {\doibase 10.1103/PhysRevA.82.011604} {\bibfield
  {journal} {\bibinfo  {journal} {Phys. Rev. A}\ }\textbf {\bibinfo {volume}
  {82}},\ \bibinfo {pages} {011604} (\bibinfo {year} {2010})}\BibitemShut
  {NoStop}%
\bibitem [{\citenamefont {Al-Hassanieh}\ \emph {et~al.}(2008)\citenamefont
  {Al-Hassanieh}, \citenamefont {Reboredo}, \citenamefont {Feiguin},
  \citenamefont {Gonzalez},\ and\ \citenamefont
  {Dagotto}}]{PhysRevLett.100.166403}%
  \BibitemOpen
  \bibfield  {author} {\bibinfo {author} {\bibfnamefont {K.~A.}\ \bibnamefont
  {Al-Hassanieh}}, \bibinfo {author} {\bibfnamefont {F.~A.}\ \bibnamefont
  {Reboredo}}, \bibinfo {author} {\bibfnamefont {A.~E.}\ \bibnamefont
  {Feiguin}}, \bibinfo {author} {\bibfnamefont {I.}~\bibnamefont {Gonzalez}}, \
  and\ \bibinfo {author} {\bibfnamefont {E.}~\bibnamefont {Dagotto}},\ }\href
  {\doibase 10.1103/PhysRevLett.100.166403} {\bibfield  {journal} {\bibinfo
  {journal} {Phys. Rev. Lett.}\ }\textbf {\bibinfo {volume} {100}},\ \bibinfo
  {pages} {166403} (\bibinfo {year} {2008})}\BibitemShut {NoStop}%
\bibitem [{\citenamefont {Mierzejewski}\ \emph
  {et~al.}(2011{\natexlab{a}})\citenamefont {Mierzejewski}, \citenamefont
  {Bon{\v{c}}a},\ and\ \citenamefont
  {Prelov{\v{s}}ek}}]{PhysRevLett.107.126601}%
  \BibitemOpen
  \bibfield  {author} {\bibinfo {author} {\bibfnamefont {M.}~\bibnamefont
  {Mierzejewski}}, \bibinfo {author} {\bibfnamefont {J.}~\bibnamefont
  {Bon{\v{c}}a}}, \ and\ \bibinfo {author} {\bibfnamefont {P.}~\bibnamefont
  {Prelov{\v{s}}ek}},\ }\href {\doibase 10.1103/PhysRevLett.107.126601}
  {\bibfield  {journal} {\bibinfo  {journal} {Phys. Rev. Lett.}\ }\textbf
  {\bibinfo {volume} {107}},\ \bibinfo {pages} {126601} (\bibinfo {year}
  {2011}{\natexlab{a}})}\BibitemShut {NoStop}%
\bibitem [{\citenamefont {Mierzejewski}\ \emph
  {et~al.}(2011{\natexlab{b}})\citenamefont {Mierzejewski}, \citenamefont
  {Vidmar}, \citenamefont {Bon{\v{c}}a},\ and\ \citenamefont
  {Prelov{\v{s}}ek}}]{PhysRevLett.106.196401}%
  \BibitemOpen
  \bibfield  {author} {\bibinfo {author} {\bibfnamefont {M.}~\bibnamefont
  {Mierzejewski}}, \bibinfo {author} {\bibfnamefont {L.}~\bibnamefont
  {Vidmar}}, \bibinfo {author} {\bibfnamefont {J.}~\bibnamefont {Bon{\v{c}}a}},
  \ and\ \bibinfo {author} {\bibfnamefont {P.}~\bibnamefont
  {Prelov{\v{s}}ek}},\ }\href {\doibase 10.1103/PhysRevLett.106.196401}
  {\bibfield  {journal} {\bibinfo  {journal} {Phys. Rev. Lett.}\ }\textbf
  {\bibinfo {volume} {106}},\ \bibinfo {pages} {196401} (\bibinfo {year}
  {2011}{\natexlab{b}})}\BibitemShut {NoStop}%
\bibitem [{\citenamefont {Eckstein}\ and\ \citenamefont
  {Werner}(2011)}]{eckstein_damping_2011}%
  \BibitemOpen
  \bibfield  {author} {\bibinfo {author} {\bibfnamefont {M.}~\bibnamefont
  {Eckstein}}\ and\ \bibinfo {author} {\bibfnamefont {P.}~\bibnamefont
  {Werner}},\ }\href {\doibase 10.1103/PhysRevLett.107.186406} {\bibfield
  {journal} {\bibinfo  {journal} {Phys. Rev. Lett.}\ }\textbf {\bibinfo
  {volume} {107}},\ \bibinfo {pages} {186406} (\bibinfo {year}
  {2011})}\BibitemShut {NoStop}%
\bibitem [{\citenamefont {Eckstein}\ \emph
  {et~al.}(2010{\natexlab{b}})\citenamefont {Eckstein}, \citenamefont {Oka},\
  and\ \citenamefont {Werner}}]{PhysRevLett.105.146404}%
  \BibitemOpen
  \bibfield  {author} {\bibinfo {author} {\bibfnamefont {M.}~\bibnamefont
  {Eckstein}}, \bibinfo {author} {\bibfnamefont {T.}~\bibnamefont {Oka}}, \
  and\ \bibinfo {author} {\bibfnamefont {P.}~\bibnamefont {Werner}},\ }\href
  {\doibase 10.1103/PhysRevLett.105.146404} {\bibfield  {journal} {\bibinfo
  {journal} {Phys. Rev. Lett.}\ }\textbf {\bibinfo {volume} {105}},\ \bibinfo
  {pages} {146404} (\bibinfo {year} {2010}{\natexlab{b}})}\BibitemShut
  {NoStop}%
\bibitem [{\citenamefont {Lu}\ \emph {et~al.}(2012{\natexlab{a}})\citenamefont
  {Lu}, \citenamefont {Sota}, \citenamefont {Matsueda}, \citenamefont
  {Bon{\v{c}}a},\ and\ \citenamefont {Tohyama}}]{PhysRevLett.109.197401}%
  \BibitemOpen
  \bibfield  {author} {\bibinfo {author} {\bibfnamefont {H.}~\bibnamefont
  {Lu}}, \bibinfo {author} {\bibfnamefont {S.}~\bibnamefont {Sota}}, \bibinfo
  {author} {\bibfnamefont {H.}~\bibnamefont {Matsueda}}, \bibinfo {author}
  {\bibfnamefont {J.}~\bibnamefont {Bon{\v{c}}a}}, \ and\ \bibinfo {author}
  {\bibfnamefont {T.}~\bibnamefont {Tohyama}},\ }\href {\doibase
  10.1103/PhysRevLett.109.197401} {\bibfield  {journal} {\bibinfo  {journal}
  {Phys. Rev. Lett.}\ }\textbf {\bibinfo {volume} {109}},\ \bibinfo {pages}
  {197401} (\bibinfo {year} {2012}{\natexlab{a}})}\BibitemShut {NoStop}%
\bibitem [{\citenamefont {Takahashi}\ \emph {et~al.}(2008)\citenamefont
  {Takahashi}, \citenamefont {Itoh},\ and\ \citenamefont
  {Aihara}}]{takahashi_photoinduced_2008}%
  \BibitemOpen
  \bibfield  {author} {\bibinfo {author} {\bibfnamefont {A.}~\bibnamefont
  {Takahashi}}, \bibinfo {author} {\bibfnamefont {H.}~\bibnamefont {Itoh}}, \
  and\ \bibinfo {author} {\bibfnamefont {M.}~\bibnamefont {Aihara}},\ }\href
  {\doibase 10.1103/PhysRevB.77.205105} {\bibfield  {journal} {\bibinfo
  {journal} {Phys. Rev. B}\ }\textbf {\bibinfo {volume} {77}},\ \bibinfo
  {pages} {205105} (\bibinfo {year} {2008})}\BibitemShut {NoStop}%
\bibitem [{\citenamefont {Mierzejewski}\ \emph {et~al.}(2010)\citenamefont
  {Mierzejewski}, \citenamefont {Luczka},\ and\ \citenamefont
  {Dajka}}]{mierzejewski_current_2010}%
  \BibitemOpen
  \bibfield  {author} {\bibinfo {author} {\bibfnamefont {M.}~\bibnamefont
  {Mierzejewski}}, \bibinfo {author} {\bibfnamefont {J.}~\bibnamefont
  {Luczka}}, \ and\ \bibinfo {author} {\bibfnamefont {J.}~\bibnamefont
  {Dajka}},\ }\href {\doibase 10.1088/0953-8984/22/24/245301} {\bibfield
  {journal} {\bibinfo  {journal} {J. Phys.: Condens. Matter}\ }\textbf
  {\bibinfo {volume} {22}},\ \bibinfo {pages} {245301} (\bibinfo {year}
  {2010})}\BibitemShut {NoStop}%
\bibitem [{\citenamefont {Oka}\ \emph {et~al.}(2003)\citenamefont {Oka},
  \citenamefont {Arita},\ and\ \citenamefont {Aoki}}]{oka_breakdown_2003}%
  \BibitemOpen
  \bibfield  {author} {\bibinfo {author} {\bibfnamefont {T.}~\bibnamefont
  {Oka}}, \bibinfo {author} {\bibfnamefont {R.}~\bibnamefont {Arita}}, \ and\
  \bibinfo {author} {\bibfnamefont {H.}~\bibnamefont {Aoki}},\ }\href {\doibase
  10.1103/PhysRevLett.91.066406} {\bibfield  {journal} {\bibinfo  {journal}
  {Phys. Rev. Lett.}\ }\textbf {\bibinfo {volume} {91}},\ \bibinfo {pages}
  {066406} (\bibinfo {year} {2003})}\BibitemShut {NoStop}%
\bibitem [{\citenamefont {Oka}\ and\ \citenamefont
  {Aoki}(2005)}]{PhysRevLett.95.137601}%
  \BibitemOpen
  \bibfield  {author} {\bibinfo {author} {\bibfnamefont {T.}~\bibnamefont
  {Oka}}\ and\ \bibinfo {author} {\bibfnamefont {H.}~\bibnamefont {Aoki}},\
  }\href {\doibase 10.1103/PhysRevLett.95.137601} {\bibfield  {journal}
  {\bibinfo  {journal} {Phys. Rev. Lett.}\ }\textbf {\bibinfo {volume} {95}},\
  \bibinfo {pages} {137601} (\bibinfo {year} {2005})}\BibitemShut {NoStop}%
\bibitem [{\citenamefont {Landau}(1932)}]{Landau}%
  \BibitemOpen
  \bibfield  {author} {\bibinfo {author} {\bibfnamefont {L.~D.}\ \bibnamefont
  {Landau}},\ }\href@noop {} {\bibfield  {journal} {\bibinfo  {journal} {Phys.
  Z. Owjetunion}\ }\textbf {\bibinfo {volume} {2}},\ \bibinfo {pages} {46}
  (\bibinfo {year} {1932})}\BibitemShut {NoStop}%
\bibitem [{\citenamefont {Zener}(1932)}]{Zener}%
  \BibitemOpen
  \bibfield  {author} {\bibinfo {author} {\bibfnamefont {C.}~\bibnamefont
  {Zener}},\ }\href@noop {} {\bibfield  {journal} {\bibinfo  {journal} {Proc R.
  Soc. A}\ }\textbf {\bibinfo {volume} {137}},\ \bibinfo {pages} {696}
  (\bibinfo {year} {1932})}\BibitemShut {NoStop}%
\bibitem [{\citenamefont {Lenar{\v{c}}i{\v{c}}}\ and\ \citenamefont
  {Prelov{\v{s}}ek}(2012)}]{PhysRevLett.108.196401}%
  \BibitemOpen
  \bibfield  {author} {\bibinfo {author} {\bibfnamefont {Z.}~\bibnamefont
  {Lenar{\v{c}}i{\v{c}}}}\ and\ \bibinfo {author} {\bibfnamefont
  {P.}~\bibnamefont {Prelov{\v{s}}ek}},\ }\href {\doibase
  10.1103/PhysRevLett.108.196401} {\bibfield  {journal} {\bibinfo  {journal}
  {Phys. Rev. Lett.}\ }\textbf {\bibinfo {volume} {108}},\ \bibinfo {pages}
  {196401} (\bibinfo {year} {2012})}\BibitemShut {NoStop}%
\bibitem [{\citenamefont {You}\ \emph {et~al.}(2007)\citenamefont {You},
  \citenamefont {Li},\ and\ \citenamefont {Gu}}]{you_fidelity_2007}%
  \BibitemOpen
  \bibfield  {author} {\bibinfo {author} {\bibfnamefont {W.-L.}\ \bibnamefont
  {You}}, \bibinfo {author} {\bibfnamefont {Y.-W.}\ \bibnamefont {Li}}, \ and\
  \bibinfo {author} {\bibfnamefont {S.-J.}\ \bibnamefont {Gu}},\ }\href
  {http://link.aps.org/doi/10.1103/PhysRevE.76.022101} {\bibfield  {journal}
  {\bibinfo  {journal} {Phys. Rev. E}\ }\textbf {\bibinfo {volume} {76}},\
  \bibinfo {pages} {022101} (\bibinfo {year} {2007})}\BibitemShut {NoStop}%
\bibitem [{\citenamefont {Feldmann}\ \emph {et~al.}(1992)\citenamefont
  {Feldmann}, \citenamefont {Leo}, \citenamefont {Shah}, \citenamefont
  {Miller}, \citenamefont {Cunningham}, \citenamefont {Meier}, \citenamefont
  {von Plessen}, \citenamefont {Schulze}, \citenamefont {Thomas},\ and\
  \citenamefont {Schmitt-Rink}}]{PhysRevB.46.7252}%
  \BibitemOpen
  \bibfield  {author} {\bibinfo {author} {\bibfnamefont {J.}~\bibnamefont
  {Feldmann}}, \bibinfo {author} {\bibfnamefont {K.}~\bibnamefont {Leo}},
  \bibinfo {author} {\bibfnamefont {J.}~\bibnamefont {Shah}}, \bibinfo {author}
  {\bibfnamefont {D.~A.~B.}\ \bibnamefont {Miller}}, \bibinfo {author}
  {\bibfnamefont {J.~E.}\ \bibnamefont {Cunningham}}, \bibinfo {author}
  {\bibfnamefont {T.}~\bibnamefont {Meier}}, \bibinfo {author} {\bibfnamefont
  {G.}~\bibnamefont {von Plessen}}, \bibinfo {author} {\bibfnamefont
  {A.}~\bibnamefont {Schulze}}, \bibinfo {author} {\bibfnamefont
  {P.}~\bibnamefont {Thomas}}, \ and\ \bibinfo {author} {\bibfnamefont
  {S.}~\bibnamefont {Schmitt-Rink}},\ }\href {\doibase
  10.1103/PhysRevB.46.7252} {\bibfield  {journal} {\bibinfo  {journal} {Phys.
  Rev. B}\ }\textbf {\bibinfo {volume} {46}},\ \bibinfo {pages} {7252}
  (\bibinfo {year} {1992})}\BibitemShut {NoStop}%
\bibitem [{\citenamefont {Leo}\ \emph {et~al.}(1992)\citenamefont {Leo},
  \citenamefont {Bolivar}, \citenamefont {Bruggemann}, \citenamefont
  {Schwedler},\ and\ \citenamefont {Kohler}}]{Leo1992943}%
  \BibitemOpen
  \bibfield  {author} {\bibinfo {author} {\bibfnamefont {K.}~\bibnamefont
  {Leo}}, \bibinfo {author} {\bibfnamefont {P.~H.}\ \bibnamefont {Bolivar}},
  \bibinfo {author} {\bibfnamefont {F.}~\bibnamefont {Bruggemann}}, \bibinfo
  {author} {\bibfnamefont {R.}~\bibnamefont {Schwedler}}, \ and\ \bibinfo
  {author} {\bibfnamefont {K.}~\bibnamefont {Kohler}},\ }\href {\doibase
  http://dx.doi.org/10.1016/0038-1098(92)90798-E} {\bibfield  {journal}
  {\bibinfo  {journal} {Solid State Communications}\ }\textbf {\bibinfo
  {volume} {84}},\ \bibinfo {pages} {943 } (\bibinfo {year}
  {1992})}\BibitemShut {NoStop}%
\bibitem [{\citenamefont {Waschke}\ \emph {et~al.}(1993)\citenamefont
  {Waschke}, \citenamefont {Roskos}, \citenamefont {Schwedler}, \citenamefont
  {Leo}, \citenamefont {Kurz},\ and\ \citenamefont
  {Kohler}}]{PhysRevLett.70.3319}%
  \BibitemOpen
  \bibfield  {author} {\bibinfo {author} {\bibfnamefont {C.}~\bibnamefont
  {Waschke}}, \bibinfo {author} {\bibfnamefont {H.~G.}\ \bibnamefont {Roskos}},
  \bibinfo {author} {\bibfnamefont {R.}~\bibnamefont {Schwedler}}, \bibinfo
  {author} {\bibfnamefont {K.}~\bibnamefont {Leo}}, \bibinfo {author}
  {\bibfnamefont {H.}~\bibnamefont {Kurz}}, \ and\ \bibinfo {author}
  {\bibfnamefont {K.}~\bibnamefont {Kohler}},\ }\href {\doibase
  10.1103/PhysRevLett.70.3319} {\bibfield  {journal} {\bibinfo  {journal}
  {Phys. Rev. Lett.}\ }\textbf {\bibinfo {volume} {70}},\ \bibinfo {pages}
  {3319} (\bibinfo {year} {1993})}\BibitemShut {NoStop}%
\bibitem [{\citenamefont {Voisin}\ \emph {et~al.}(1988)\citenamefont {Voisin},
  \citenamefont {Bleuse}, \citenamefont {Bouche}, \citenamefont {Gaillard},
  \citenamefont {Alibert},\ and\ \citenamefont
  {Regreny}}]{PhysRevLett.61.1639}%
  \BibitemOpen
  \bibfield  {author} {\bibinfo {author} {\bibfnamefont {P.}~\bibnamefont
  {Voisin}}, \bibinfo {author} {\bibfnamefont {J.}~\bibnamefont {Bleuse}},
  \bibinfo {author} {\bibfnamefont {C.}~\bibnamefont {Bouche}}, \bibinfo
  {author} {\bibfnamefont {S.}~\bibnamefont {Gaillard}}, \bibinfo {author}
  {\bibfnamefont {C.}~\bibnamefont {Alibert}}, \ and\ \bibinfo {author}
  {\bibfnamefont {A.}~\bibnamefont {Regreny}},\ }\href {\doibase
  10.1103/PhysRevLett.61.1639} {\bibfield  {journal} {\bibinfo  {journal}
  {Phys. Rev. Lett.}\ }\textbf {\bibinfo {volume} {61}},\ \bibinfo {pages}
  {1639} (\bibinfo {year} {1988})}\BibitemShut {NoStop}%
\bibitem [{\citenamefont {Mierzejewski}\ and\ \citenamefont
  {Prelov{\v{s}}ek}(2010)}]{PhysRevLett.105.186405}%
  \BibitemOpen
  \bibfield  {author} {\bibinfo {author} {\bibfnamefont {M.}~\bibnamefont
  {Mierzejewski}}\ and\ \bibinfo {author} {\bibfnamefont {P.}~\bibnamefont
  {Prelov{\v{s}}ek}},\ }\href {\doibase 10.1103/PhysRevLett.105.186405}
  {\bibfield  {journal} {\bibinfo  {journal} {Phys. Rev. Lett.}\ }\textbf
  {\bibinfo {volume} {105}},\ \bibinfo {pages} {186405} (\bibinfo {year}
  {2010})}\BibitemShut {NoStop}%
\bibitem [{\citenamefont {Vidmar}\ \emph {et~al.}(2011)\citenamefont {Vidmar},
  \citenamefont {Bon{\v{c}}a}, \citenamefont {Mierzejewski}, \citenamefont
  {Prelov{\v{s}}ek},\ and\ \citenamefont
  {Trugman}}]{vidmar_nonequilibrium_2011}%
  \BibitemOpen
  \bibfield  {author} {\bibinfo {author} {\bibfnamefont {L.}~\bibnamefont
  {Vidmar}}, \bibinfo {author} {\bibfnamefont {J.}~\bibnamefont {Bon{\v{c}}a}},
  \bibinfo {author} {\bibfnamefont {M.}~\bibnamefont {Mierzejewski}}, \bibinfo
  {author} {\bibfnamefont {P.}~\bibnamefont {Prelov{\v{s}}ek}}, \ and\ \bibinfo
  {author} {\bibfnamefont {S.~A.}\ \bibnamefont {Trugman}},\ }\href {\doibase
  10.1103/PhysRevB.83.134301} {\bibfield  {journal} {\bibinfo  {journal} {Phys.
  Rev. B}\ }\textbf {\bibinfo {volume} {83}},\ \bibinfo {pages} {134301}
  (\bibinfo {year} {2011})}\BibitemShut {NoStop}%
\bibitem [{\citenamefont {Tomka}\ \emph {et~al.}(2012)\citenamefont {Tomka},
  \citenamefont {Polkovnikov},\ and\ \citenamefont
  {Gritsev}}]{PhysRevLett.108.080404}%
  \BibitemOpen
  \bibfield  {author} {\bibinfo {author} {\bibfnamefont {M.}~\bibnamefont
  {Tomka}}, \bibinfo {author} {\bibfnamefont {A.}~\bibnamefont {Polkovnikov}},
  \ and\ \bibinfo {author} {\bibfnamefont {V.}~\bibnamefont {Gritsev}},\ }\href
  {\doibase 10.1103/PhysRevLett.108.080404} {\bibfield  {journal} {\bibinfo
  {journal} {Phys. Rev. Lett.}\ }\textbf {\bibinfo {volume} {108}},\ \bibinfo
  {pages} {080404} (\bibinfo {year} {2012})}\BibitemShut {NoStop}%
\bibitem [{\citenamefont {Zanardi}\ and\ \citenamefont
  {Paunkovi{\'{c}}}(2006)}]{PhysRevE.74.031123}%
  \BibitemOpen
  \bibfield  {author} {\bibinfo {author} {\bibfnamefont {P.}~\bibnamefont
  {Zanardi}}\ and\ \bibinfo {author} {\bibfnamefont {N.}~\bibnamefont
  {Paunkovi{\'{c}}}},\ }\href {\doibase 10.1103/PhysRevE.74.031123} {\bibfield
  {journal} {\bibinfo  {journal} {Phys. Rev. E}\ }\textbf {\bibinfo {volume}
  {74}},\ \bibinfo {pages} {031123} (\bibinfo {year} {2006})}\BibitemShut
  {NoStop}%
\bibitem [{\citenamefont {Greschner}\ \emph {et~al.}(2013)\citenamefont
  {Greschner}, \citenamefont {Kolezhuk},\ and\ \citenamefont
  {Vekua}}]{PhysRevB.88.195101}%
  \BibitemOpen
  \bibfield  {author} {\bibinfo {author} {\bibfnamefont {S.}~\bibnamefont
  {Greschner}}, \bibinfo {author} {\bibfnamefont {A.~K.}\ \bibnamefont
  {Kolezhuk}}, \ and\ \bibinfo {author} {\bibfnamefont {T.}~\bibnamefont
  {Vekua}},\ }\href {\doibase 10.1103/PhysRevB.88.195101} {\bibfield  {journal}
  {\bibinfo  {journal} {Phys. Rev. B}\ }\textbf {\bibinfo {volume} {88}},\
  \bibinfo {pages} {195101} (\bibinfo {year} {2013})}\BibitemShut {NoStop}%
\bibitem [{\citenamefont {Ejima}\ and\ \citenamefont
  {Nishimoto}(2007)}]{PhysRevLett.99.216403}%
  \BibitemOpen
  \bibfield  {author} {\bibinfo {author} {\bibfnamefont {S.}~\bibnamefont
  {Ejima}}\ and\ \bibinfo {author} {\bibfnamefont {S.}~\bibnamefont
  {Nishimoto}},\ }\href {\doibase 10.1103/PhysRevLett.99.216403} {\bibfield
  {journal} {\bibinfo  {journal} {Phys. Rev. Lett.}\ }\textbf {\bibinfo
  {volume} {99}},\ \bibinfo {pages} {216403} (\bibinfo {year}
  {2007})}\BibitemShut {NoStop}%
\bibitem [{\citenamefont {Oka}\ \emph {et~al.}(2005)\citenamefont {Oka},
  \citenamefont {Konno}, \citenamefont {Arita},\ and\ \citenamefont
  {Aoki}}]{PhysRevLett.94.100602}%
  \BibitemOpen
  \bibfield  {author} {\bibinfo {author} {\bibfnamefont {T.}~\bibnamefont
  {Oka}}, \bibinfo {author} {\bibfnamefont {N.}~\bibnamefont {Konno}}, \bibinfo
  {author} {\bibfnamefont {R.}~\bibnamefont {Arita}}, \ and\ \bibinfo {author}
  {\bibfnamefont {H.}~\bibnamefont {Aoki}},\ }\href {\doibase
  10.1103/PhysRevLett.94.100602} {\bibfield  {journal} {\bibinfo  {journal}
  {Phys. Rev. Lett.}\ }\textbf {\bibinfo {volume} {94}},\ \bibinfo {pages}
  {100602} (\bibinfo {year} {2005})}\BibitemShut {NoStop}%
\bibitem [{\citenamefont {Poilblanc}\ \emph {et~al.}(1993)\citenamefont
  {Poilblanc}, \citenamefont {Ziman}, \citenamefont {Bellissard}, \citenamefont
  {Mila},\ and\ \citenamefont {Montambaux}}]{poilblanc_poisson_1993}%
  \BibitemOpen
  \bibfield  {author} {\bibinfo {author} {\bibfnamefont {D.}~\bibnamefont
  {Poilblanc}}, \bibinfo {author} {\bibfnamefont {T.}~\bibnamefont {Ziman}},
  \bibinfo {author} {\bibfnamefont {J.}~\bibnamefont {Bellissard}}, \bibinfo
  {author} {\bibfnamefont {F.}~\bibnamefont {Mila}}, \ and\ \bibinfo {author}
  {\bibfnamefont {G.}~\bibnamefont {Montambaux}},\ }\href {\doibase
  10.1209/0295-5075/22/7/010} {\bibfield  {journal} {\bibinfo  {journal}
  {{Euro. Phys. Lett.}}\ }\textbf {\bibinfo {volume} {22}},\ \bibinfo {pages}
  {537} (\bibinfo {year} {1993})}\BibitemShut {NoStop}%
\bibitem [{\citenamefont {Lu}\ \emph {et~al.}(2012{\natexlab{b}})\citenamefont
  {Lu}, \citenamefont {Sota}, \citenamefont {Matsueda}, \citenamefont
  {Bon{\v{c}}a},\ and\ \citenamefont {Tohyama}}]{lu_enhanced_2012}%
  \BibitemOpen
  \bibfield  {author} {\bibinfo {author} {\bibfnamefont {H.}~\bibnamefont
  {Lu}}, \bibinfo {author} {\bibfnamefont {S.}~\bibnamefont {Sota}}, \bibinfo
  {author} {\bibfnamefont {H.}~\bibnamefont {Matsueda}}, \bibinfo {author}
  {\bibfnamefont {J.}~\bibnamefont {Bon{\v{c}}a}}, \ and\ \bibinfo {author}
  {\bibfnamefont {T.}~\bibnamefont {Tohyama}},\ }\href {\doibase
  10.1103/PhysRevLett.109.197401} {\bibfield  {journal} {\bibinfo  {journal}
  {Phys. Rev. Lett.}\ }\textbf {\bibinfo {volume} {109}},\ \bibinfo {pages}
  {197401} (\bibinfo {year} {2012}{\natexlab{b}})}\BibitemShut {NoStop}%
\end{thebibliography}
%merlin.mbs apsrev4-1.bst 2010-07-25 4.21a (PWD, AO, DPC) hacked
%Control: key (0)
%Control: author (72) initials jnrlst
%Control: editor formatted (1) identically to author
%Control: production of article title (-1) disabled
%Control: page (0) single
%Control: year (1) truncated
%Control: production of eprint (0) enabled
%

\end{document}